\documentclass[a4paper,journal]{IEEEtran}
\usepackage{cite}
\usepackage{amsmath}
\usepackage{array}
\usepackage{amssymb}

\usepackage[pdftex]{graphicx}
\usepackage{caption}
\usepackage{subcaption}

\usepackage{float}
\usepackage{lineno,hyperref}
\usepackage{verbatim}

\usepackage{algorithmicx}
\usepackage[ruled]{algorithm}
\usepackage{algpseudocode}

\usepackage{soul}
\usepackage{graphicx}
\usepackage{tikz}
\usetikzlibrary{trees,positioning,shapes,shadows,arrows.meta}
\usepackage[edges]{forest}

\begin{document}

\title{Geometric Deep Learning for Computer-Aided Design: A Survey}

\author{\IEEEauthorblockN{Negar Heidari and Alexandros Iosifidis}\\
\IEEEauthorblockA{Department of Electrical and Computer Engineering, Aarhus University, Denmark\\
Emails: \{negar.heidari, ai\}@ece.au.dk}}
\maketitle

\begin{abstract}
Geometric Deep Learning techniques have become a transformative force in the field of Computer-Aided Design (CAD), and have the potential to revolutionize how designers and engineers approach and enhance the design process. By harnessing the power of machine learning-based methods, CAD designers can optimize their workflows, save time and effort while making better informed decisions, and create designs that are both innovative and practical. The ability to process the CAD designs represented by geometric data and to analyze their encoded features enables the identification of similarities among diverse CAD models, the proposition of alternative designs and enhancements, and even the generation of novel design alternatives.
This survey offers a comprehensive overview of learning-based methods in computer-aided design across various categories, including similarity analysis and retrieval, 2D and 3D CAD model synthesis, and CAD generation from point clouds, and single/multi-view images. Additionally, it provides a complete list of benchmark datasets and their characteristics, along with open-source codes that have propelled research in this domain. The final discussion delves into the challenges prevalent in this field, followed by potential future research directions in this rapidly evolving field.
\end{abstract}


\IEEEpeerreviewmaketitle


\section{Introduction}

Deep Learning has revolutionized data processing across different domains spanning from grid-structured data in Euclidean spaces, such as image, video, text, to graph-structured data with complex relationships. 
Geometric Deep Learning (GDL) encompasses a wide array of neural network architectures ranging from Convolutional Neural Networks (CNNs), Graph Neural Networks (GNNs), Recurrent Neural Networks (RNNs), to Transformer Networks, all of which encode a geometric understanding of the data such as symmetry and invariance as an inductive bias in their learning process \cite{bronstein2021geometric}. 
Over the past decade, GDL methods, including CNNs, GNNs, RNNs, and Transformer Networks, have made remarkable strides in diverse tasks coming from different applications, including computer vision, natural language processing, computer graphics, and bioinformatics. However, the application of GDL methods on complex parametric Computer-Aided Design (CAD) data is rarely studied. 
Boundary Representation (B-Rep), which is a fundamental data format for CAD models encoding a high level of parametric details in CAD models, can be used to learn the intricate geometric features of such models.

Although GDL methods have seen considerable success in analyzing 3D shapes in mesh \cite{kalogerakis20173d, feng2019meshnet}, voxel \cite{maturana2015voxnet, wu20153d, wang2019normalnet}, and point cloud \cite{le2018pointgrid, qi2017pointnet} data formats, extracting feature representations directly from B-Rep data poses challenges. Converting B-Rep data to conventional formats like triangle meshes is not only computationally expensive but also leads to information loss \cite{jayaraman2021uv}. Hence, learning feature representations directly from B-Rep data becomes imperative, ensuring an efficient capture of the most representative geometric features in CAD models for subsequent analyses without the burden of extensive computation and memory usage.

Recently, there has been a great research interest in leveraging GDL methods for learning the structure of CAD models and for facilitating the design process in different aspects. 
While conventional machine learning and deep learning approaches have been explored for CAD classification and clustering tasks \cite{krahe2020deep, krahe2022ai, machalica2019cad}, these methods predominantly focus on learning feature representations based on physical features like volume, bounding box, area, density, mass, principal axes, or directly work on other data formats for 3D data, such as point cloud, without considering B-Rep. Accordingly, they often fall short in capturing the concise geometric properties embedded in CAD models.

As GDL methods continue to evolve, they are expected to play a pivotal role in shaping the future of CAD design across industries.
In recent years, more advanced GDL methods such as GNNs and (Graph) Transformer Networks have shown great potential in learning the complex geometric features embedded in CAD models, particularly from B-Rep data, irrespective of their physical features \cite{jayaraman2021uv, jones2023self, lambourne2021brepnet, krahe2020deep, solidgen23, zhou2023cadparser}. 
These learned feature representations serve various purposes, such as reconstructing CAD models \cite{wu2021deepcad, solidgen23, zhou2023cadparser}, determining joints between CAD solids \cite{willis2022joinable, jones2021automate}, and autocompleting unfinished CAD models \cite{seff2020sketchgraphs, willis2021engineering, li2022free2cad, SeffZRA22}. The overarching objective of these learning-based approaches is to elevate the level of automation in the CAD design process. By alleviating the need for experts to perform repetitive and time-consuming tasks, such as sketching and drafting, these methods have the potential to empower designers to focus on the more creative facets of the design process.
By analyzing historical design data, extracting geometric features, and discerning valuable insights like similarities between CAD models, these methods can facilitate the reuse of CAD models in new products. This not only saves time and resources by preventing redundant designs but also aids designers in customization by generating design alternatives based on specific parameters and objectives. 
Furthermore, there is a growing demand for CAD tools that can reverse engineer models and generate diverse design options based on concise parameters derived from preliminary concept sketches. GDL-based methodologies have made some progress in meeting this demand \cite{li2020sketch2cad, li2022free2cad}.
However, the introduction of machine learning-based methods in this area is not without challenges. A major challenge is the scarcity of annotated CAD datasets in B-Rep format. Unlike other data formats such as images, videos, or text, collecting CAD models is intricate and time-consuming, typically requiring the expertise of skilled engineers. Moreover, these datasets often remain proprietary and inaccessible to the public. On the other hand, annotating CAD datasets, especially for mechanical objects, demands substantial domain knowledge. Given that training deep learning methods without large-scale annotated datasets is impractical, recent research efforts have also made valuable contributions by providing benchmark CAD datasets.

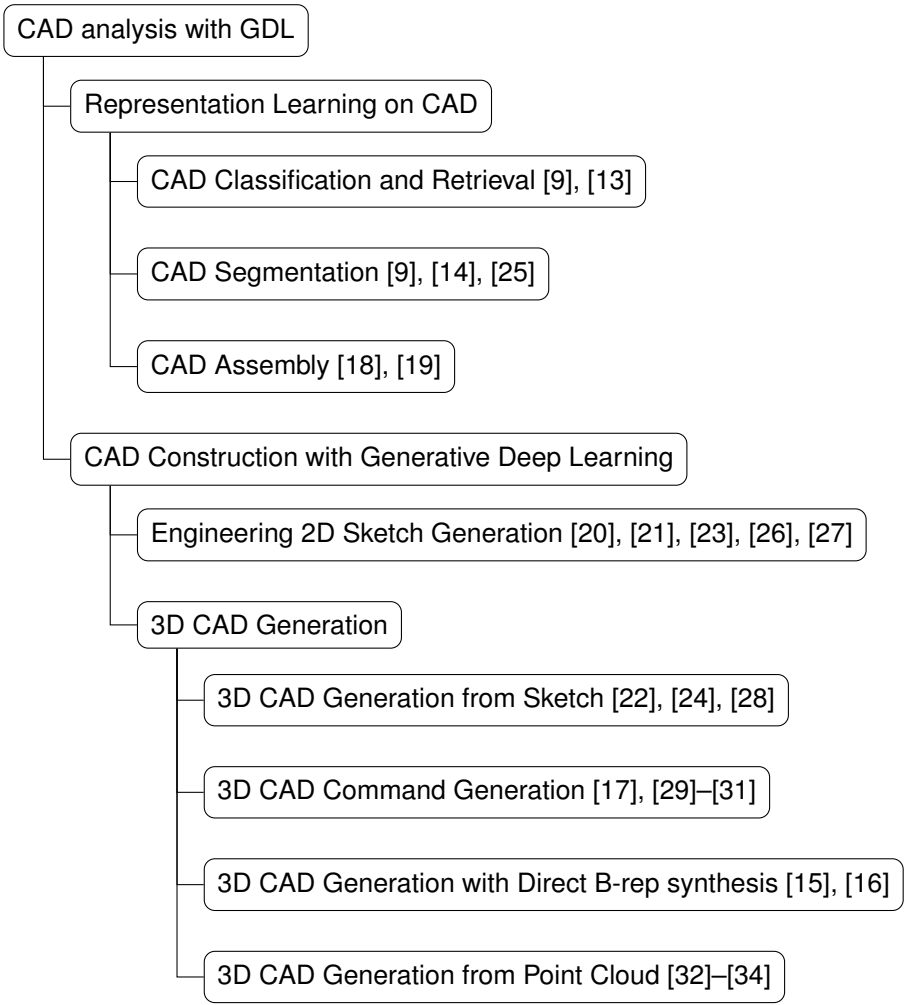
\begin{figure*}[ht]
\begin{forest}
  for tree={
    font=\sffamily,
    grow'=0,
    child anchor=west,
    parent anchor=south,
    anchor=west,
    calign=first,
    edge path={
      \noexpand\path [draw, \forestoption{edge}] (!u.south west) +(15pt,0) |- (.child anchor) \forestoption{edge label};
    },
    before typesetting nodes={
      if n=1
        {insert before={[,phantom]}}
        {}
    },
    fit=band,
    before computing xy={l=25pt},
    s sep=15pt, 
    draw,
    inner sep=5pt,
    rounded corners,
  }
  [CAD analysis with GDL
    [Representation Learning on CAD
        [CAD Classification and Retrieval \cite{jayaraman2021uv, jones2023self}]
        [CAD Segmentation \cite{jayaraman2021uv, lambourne2021brepnet, dupont2022cadops}]
        [CAD Assembly \cite{jones2021automate, willis2022joinable}]
    ]
    [CAD Construction with Generative Deep Learning
        [Engineering 2D Sketch Generation \cite{seff2020sketchgraphs, willis2021engineering, para2021sketchgen, ganin2021computer, SeffZRA22}]
        [3D CAD Generation
            [3D CAD Generation from Sketch \cite{li2020sketch2cad, li2022free2cad, hahnlein2022cad2sketch}]
            [3D CAD Command Generation \cite{xu2021inferring, willis2021fusion, wu2021deepcad, skexgen22, zhang2025diffusion}]
            [3D CAD Generation with Direct B-rep synthesis \cite{solidgen23, zhou2023cadparser, xu2024brepgen, li2025dtgbrepgen}]
            [3D CAD Generation from Point Cloud \cite{uy2022point2cyl, lambourne2022reconstructing, guo2022complexgen, ma2024draw}]
            [3D CAD Generation from Image \cite{li2025caddreamer, chen2025cadcrafter}]
        ]
    ]
  ]
\end{forest}
\caption{Taxonomy and the structure of the most representative methods reviewed in this survey.}
\label{fig:taxonomy}
\end{figure*}

As this field is still emerging, there is plenty of room for future research and development, and several challenges need to be explored and addressed. 
Our survey, to the best of our knowledge, stands as the first comprehensive review in this rapidly evolving domain, which endeavors to provide a thorough exploration of recent advancements, challenges, and contributions in this area. 
Figure \ref{fig:taxonomy} shows the structure of the survey and highlights the most representative methods in the field reviewed in more detail. 
As we embark on this survey, our goal is to serve as a guiding resource for researchers and practitioners eager to navigate the dynamic intersection of GDL and CAD. 
The survey is tailored to provide:
\begin{itemize}
    \item A comprehensive review of the state-of-the-art GDL methods employed in the analysis of CAD data. This encompasses diverse categories, including similarity analysis, classification, and retrieval, as well as segmentation, along with the synthesis of 2D and 3D CAD models. Furthermore, the survey explores techniques for generating CAD models from alternative data representations, such as point clouds.
    \item A detailed overview of benchmark CAD datasets crucial for the advancement of research in this field, accompanied by open-source codes that have been instrumental in pushing the boundaries of achievable outcomes.
    \item An in-depth discussion about the challenges that persist in this domain followed by potential future research directions by assessing the limitations of existing methodologies. 
\end{itemize}

\section{Background}
\subsection{Computer Aided Design (CAD)}
Computer-Aided Design (CAD) is a manufacturing technology that has revolutionized the way engineers, architects, designers, and other professionals create and visualize designs \cite{groover1983cad}. CAD process involves the use of specialized software to design, modify, analyze, and optimize 2D drawings and 3D models of physical objects or systems digitally before constructing them. Engineering drawing entails the use of graphical symbols such as points, lines, curves, planes and shapes, and it essentially gives detailed description about any component in a graphical form. 
There are several CAD software options which are widely used in industry for different purposes. 

\subsubsection{CAD Tools and Python APIs}
To name some of the most popular software options, AutoCAD \cite{autocad} and Fusion 360 \cite{fusion360} (a cloud-based software) offer 2D/3D design tools for a variety of industries. SolidWorks \cite{solidworks} and CATIA \cite{catia} provide parametric modeling and simulation for the mechanical, aerospace, and automotive industries. Onshape \cite{onshape} (a cloud-based software) and Creo \cite{creo} offer 3D parametric design and simulation. 
TinkerCAD \cite{tinkercad} is a free-of-charge and web-based CAD design tool which is mostly used for beginners and educational purposes. 

For CAD parsing and development, several Python APIs are available, providing tools and libraries to work with CAD data, create 3D models, visualize CAD models, and to perform various related operations. To name some of the most popular Python APIs for CAD, PythonOCC \cite{pythonocc} offers an open-source 3D modeling and CAD library for Python. It is based on OpenCASCADE Technology (OCCT) \cite{occt} which is a powerful geometry kernel and modeling framework used for CAD. 
PythonOCC provides Python bindings to the OpenCASCADE C++ library, allowing developers to use Python for creating, manipulating, visualizing and analyzing 3D geometry and CAD models.
OCCWL \cite{occwl} has been recently released as a simple, lightweight Pythonic wrapper around PythonOCC.
CADQuery \cite{cadquery}, a Python library for scripting 3D CAD models, is also built on top of the OpenCASCADE geometry kernel.

Another powerful 3D geometric modeling kernel is Parasolid \cite{parasolid}, which provides a set of functions and methods for creating, manipulating, and visualizing and analyzing 3D solid models. 
However, unlike OpenCASCADE, Parasolid is not open-source and software developers need to license it to integrate its capabilities into CAD applications.


\subsubsection{Terminology in CAD Design}
Boundary Representation (B-Rep), serves as a fundamental modeling technique utilized in CAD to represent intricate 3D models along with their geometrical attributes. It facilitates accurate and consistent design, modification, manipulation, analysis and representation of 3D entities by characterizing surfaces, curves, points, and topological relationships between them in 3D space. 
In Table \ref{table:CADTerminology}, we introduce the basic terminology of CAD design which is essential for understanding the concepts and methodologies in this field. However, depending on the industries and different design software applications that are used for design, there might exist more specialized terms and concepts that are not covered here. For the sake of maintaining consistency throughout this paper, we will employ this terminology for all methodologies and datasets discussed in the subsequent sections.

\begin{table*}[ht!]
	\centering
	\resizebox{\textwidth}{!}{
    \begin{tabular}{p{0.2\textwidth}|p{0.8\textwidth}}
    \hline
    \textbf{Term} & \textbf{Description} \\
    \hline
    Model & 2D or 3D representation of a real-world object or system created within a CAD software.\\
    \hline
    B-Rep & Boundary Representation, a data format for geometrically describing objects by representing their topological components such as surfaces, edges, vertices, and the relationship between them. \\ 
    \hline
    Sketch & 2D drawing of an object, served as the basis for creating 3D models, made of lines, curves, and other basic geometric shapes. \\ 
    \hline
    Extrude & A CAD operation for expanding a 2D drawing along a specific dimension to create a 3D model. \\  
    \hline
    Primitive & A basic 2D/3D geometric shape which serves as backbone for more complex designs. For 2D sketches, primitives are points, lines, arcs, circles, ellipses and polygons. For 3D shapes primitives are cubes, spheres, cylinders, cones. \\ 
    \hline
    Constraint & Geometric relationship, such as coincident, tangent, perpendicular, between primitives in a 2D/3D design.\\ 
    \hline
    Point & Basic geometric entity representing a specific location in space with 3D coordinates X, Y, Z. \\
    \hline
    Line & A straight path between two points in a 3D space. \\
    \hline
    Circle & A closed curve defined by its center and radius. \\
    \hline
    Arc & A curved line, as a part of a circle or ellipse, defined by a center, start and end points. \\
    \hline
    Loop & A closed sequence of curves forming the boundary of a surface. \\ 
    \hline
    Profile & A closed loop made by a collection of curves joint together. \\
    \hline
    Face & A 2D surface bounded by a profile. \\ 
    \hline
    Edge & A piece of a curve bounded by two points, defining the boundaries of a face. \\
    \hline
    Shell & A collection of connected faces with joint boundaries and points. \\
    \hline
    Body/Part & A 3D shape created by extruding a face. \\
    \hline
    Solid & A 3D shape that is used as a building block to design an object. \\
    \hline
    Component & Building blocks of assemblies containing one or more bodies (parts). \\ 
    \hline
    Assembly & A collection of connected bodies (or parts) to create a larger object. \\
    \hline
    Joint & The connection, defined by coordinate and axis, between bodies (parts) to make an assembly. \\
    \hline
    Topology & The structure of points, curves, faces and other geometric components in a 3D model. \\
    \hline
    Wireframe & Visual representation of 3D B-Rep models, showing the structure of the object with only lines and curves. Also known as skeletal representation of 3D objects. \\
    \hline
    Rendering & The process of visualizing CAD models to simulate their real-world appearance. \\ 
    \hline
    Surface Normal & A unit vector perpendicular to a surface at a specific point. \\ \hline
    \end{tabular} }
 \caption{Terminology in CAD design. A summary of the widely used terms through different research works and CAD platforms.} 
 \label{table:CADTerminology}
\end{table*}

\begin{table*}[!t]
	\centering
	\resizebox{\linewidth}{!}{
	
    \begin{tabular}{p{0.1\textwidth}|p{0.7\textwidth}|p{0.2\textwidth}}
    \hline
    \textbf{Type} & \textbf{Description} & \textbf{Format} \\
    \hline
    STEP & Original parametric B-Rep format containing explicit description of the topology and geometry information of the CAD models. & .txt, .step, .smt \\ 
    \hline
    STL & Standard Tessellation Language, a format describing 3D surfaces of an object with triangular facets (meshes).  & .stl \\
    \hline
    Obj & A format describing 3D surface geometry using a triangular meshes. & .obj \\
    \hline
    Statistics & Statistical infromation of CAD models. & .yml, .txt \\
    \hline
    Features & A description of the properties of surfaces and curves with references to the corresponding vertices and faces of the discrete triangle mesh representation.  & .yml \\ 
    \hline
    Image & Visualization of objects produced by rendering process. & .png \\  
    \hline
    PCD & Point Cloud Data, a collection of points in the 3D space representing the structure of a 3D model. & .ASC, .PTX, .OBJ\\ 
    \hline

    \end{tabular} }
 \caption{A summary of different file formats and their corresponding description. Every CAD dataset provides at least one of these data formats to be processed for different purposes.} 
 \label{table:CADFormats}
\end{table*}

\subsection{Geometric Deep Learning (GDL)}

Geometric Deep Learning (GDL) has arisen as a special and fundamental branch of artificial intelligence (AI) that expands deep learning approaches from regular Euclidean data to complex geometric structured data \cite{bronstein2021geometric}.
While traditional deep learning methods have led to great advancements in different applications by processing regular Euclidean data structures, such as audio, text, images and videos, formed by regular 1D, 2D and 3D grids, GDL is tailored for processing and extracting intricate spatial and topological features from irregular structured data like 3D shapes, meshes, point clouds, and graphs. These irregular data structures processed by GDL can be either Euclidean or non-Euclidean, depending on the context and the geometric properties they possess.
GDL has gained significant research attention in the CAD domain for its ability in learning complex geometric features and facilitating design process for engineers and designers. 

The field of geometric deep learning for CAD has progressed through distinct stages, each addressing limitations of the previous generation of models in handling the structured, parametric, and semantic-rich nature of CAD data.
Early approaches such as MeshCNN \cite{hanocka2019meshcnn} operated on triangle meshes using edge-based convolutions to capture local surface structures. While effective for organic shapes with well-formed manifold meshes, these models struggle with the topological complexity and degeneracies typical in industrial CAD exports, such as T-vertices, non-manifold edges, and inconsistent triangulations.

To improve robustness, point-based models like PointNet \cite{qi2017pointnet} and DGCNN \cite{wang2019dynamic} were introduced, which treat 3D shapes as unordered sets of points and learn global or local geometric features without relying on mesh connectivity. These methods offer greater resilience to topological inconsistencies and mesh noise. However, they come with a critical drawback: they discard the explicit topological and parametric structure of CAD models, including analytic surface types, like planes, cylinders, design features, and constraint relationships. As such, while useful for tasks like classification or segmentation, they fall short in applications that require understanding or manipulation of design intent.

To bridge this gap, a new class of graph-based models emerged that leverage the native boundary representation (B-Rep) topology inherent to CAD data. These Graph Neural Networks (GNNs) operate on structured graphs built from B-Rep entities, faces, edges, and vertices, enabling explicit encoding of parametric surfaces, topological constraints, and hierarchical relationships between geometric features. Models such as UV-Net \cite{jayaraman2021uv}, DeepCAD \cite{wu2021deepcad} and BRepNet \cite{lambourne2021brepnet} exemplify this direction, offering better adaptability to CAD workflows and supporting tasks like feature recognition, constraint extraction, and semantic segmentation on structured geometry.

Building upon this foundation, recent Transformer-based models such as SketchGen \cite{para2021sketchgen}, CAD As Language \cite{ganin2021computer}, PolyGen \cite{nash2020polygen} and other recent generative methods reviewed in this paper as illustrated in \ref{fig:taxonomy} represent the current state of the art. These models operate directly on tokenized B-Rep elements or design operation sequences, capturing long-range dependencies and supporting learning across rich semantic and geometric contexts. They are not only capable of encoding exact geometry, but also learn design history, manufacturability constraints, and support editability and reverse engineering, aligning closely with how CAD software encodes and manipulates geometry.

This technological evolution, from mesh- and point cloud-based methods to graph- and transformer-based architectures, demonstrates a clear trajectory toward CAD-native learning, with increasing alignment to the structural, semantic, and engineering-specific properties of design data.

\subsection{Graph neural networks (GNNs)}
Graph neural networks (GNNs) \cite{scarselli2008graph, wu2020comprehensive} are one of the most popular types of GDL approaches which excel in processing graph-structured data. GNNs have made remarkable strides in diverse tasks tasks coming from different applications, such as computer vision \cite{heidari2021temporal, heidari2021pstgcn, hedegaard2023continual}, bio-chemical pattern recognition \cite{mansimov2019molecular, wang2022molecular}, and financial data analysis \cite{baltakys2023predicting}. 
GNNs offer a specialized approach for modeling complex CAD geometric structures encoded in B-Rep, as a graph which represents the topology and geometry of 3D shapes through nodes and edges. GNN models can effectively capture the topological structure, connectivity, and attributes inherent in graphs (B-Reps) and allow hierarchical analysis by considering local and global contexts, aligning with the complex structure of B-Rep models.
GNNs are also able to extract and propagate geometric features through the graph, capturing subtle aspects of shape and curvatures and leveraging data-driven insights from existing B-Rep models, facilitating tasks like solid and/or sketch segmentation \cite{lambourne2021brepnet, jayaraman2021uv, yang2021sketchgnn}, shape reconstruction and generation \cite{zhou2023cadparser, seff2020sketchgraphs}, shape analysis and retrieval \cite{jones2023self, jayaraman2021uv}. By representing B-Rep as a graph and leveraging GNN capabilities to capture its topological structure, connectivity, and geometry, designers and engineers can analyze, optimize, and create intricate 3D models with enhanced precision and efficiency. 

\begin{figure}
    \centering
    \includegraphics[width=0.65\linewidth]{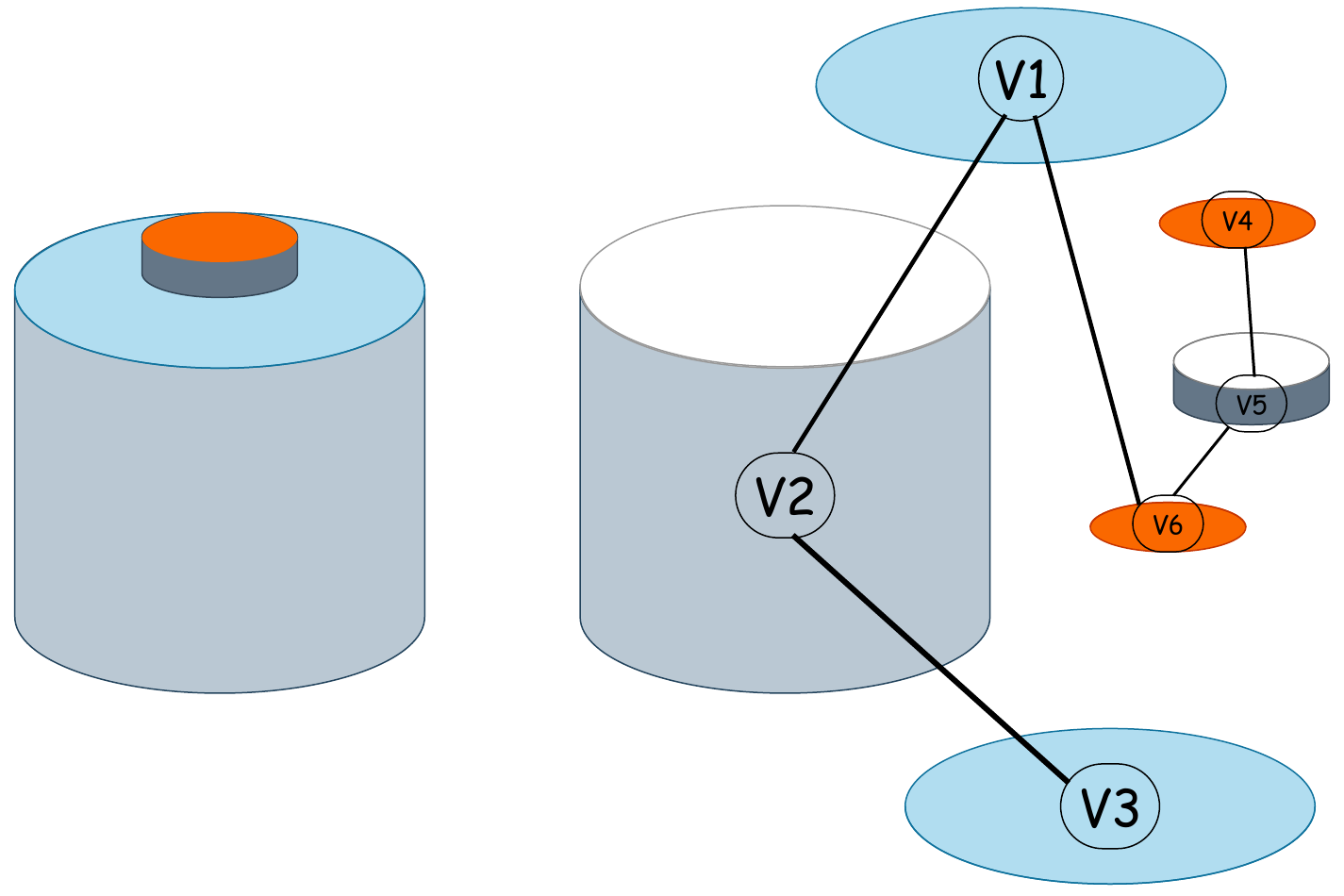}
    \caption{An example of a 3D solid represented as a graph, where the solid primitives such as curves and surfaces are represented as graph nodes, and their adjacency relationship between the solid primitives are represented as graph edges.}
    \label{fig:solidgraph}
\end{figure}

Let us present a more detailed explanation of how B-Rep models are represented as graphs. In the context of B-Rep, graph nodes can represent points, lines, and faces with associated attributes like geometric coordinates, line lengths, curvature, face area and surface normals. 
Edges in the graph represent connections between nodes. For B-Rep, these connections indicate constraints, like adjacency, tangency, or coincidence, between primitives like points, lines, faces, etc.
GNN methods embed each node's features, creating node representations that capture geometric attributes that aggregate information from neighboring nodes. In the B-Rep context, this simulates the way geometric properties flow through adjacent points, lines, curves, and faces.
Through feature aggregation, the method updates each node's representation by combining its own features with information of its neighbors. This step captures the influence of neighboring geometry.
The core operation in GNNs is the graph convolution, which is combined with a nonlinear function applied on the fused information from neighboring nodes. In a multi-layer GNN, the graph convolution operation captures multi-hop neighboring features allowing for hierarchical analysis.  
For B-Rep, this operation helps to capture the hierarchical relationships between primitives. 
To obtain a global representation of the B-Rep model, GNNs can employ global pooling, summarizing the entire graph information into a single feature vector.
Figure \ref{fig:solidgraph}, \ref{fig:sketchgraph} show an example of a 3D shape and a 2D sketch represented as graph, respectively. 

The most popular GNN architectures, i.e., Graph Convolutional Networks (GCNs) \cite{KipfW17, heidari2021pgcn, heidari2022graph} and Graph AutoEncoders (GAEs) \cite{li2018learning}, which are used for (semi-)supervised and unsupervised learning on graphs, respectively, for different graph analysis tasks such as node/graph classification and graph reconstruction/generation. 
As GNN based methods continue to evolve, they are expected to play a pivotal role in shaping the future of CAD design across industries. In the following sections, we delve into details of the current state-of-the-art methods in this area and the challenges.

\section{Datasets}
\label{sec:datasets}
Large data collections play a pivotal role in enhancing performance of deep learning models when applied in problems coming from different applications. Collecting such large datasets in regular formats such as image, video, audio, text, and their distribution through different platforms like social media has greatly accelerated the progress of deep learning in computer vision and natural language processing.  
GDL offers advantages for tasks like 3D shape analysis, shape reconstruction, and building geometric feature descriptors. Nonetheless, creating and annotating high-quality 3D geometric data needs significant level of domain knowledge, and engineering and design skills. Collecting such datasets is also challenging due to various factors such as concerns about proprietary rights and ownership, and lack of consistency and compatibility among data from available sources. 
\begin{figure}[t]
    \centering
    \includegraphics[width=1\linewidth]{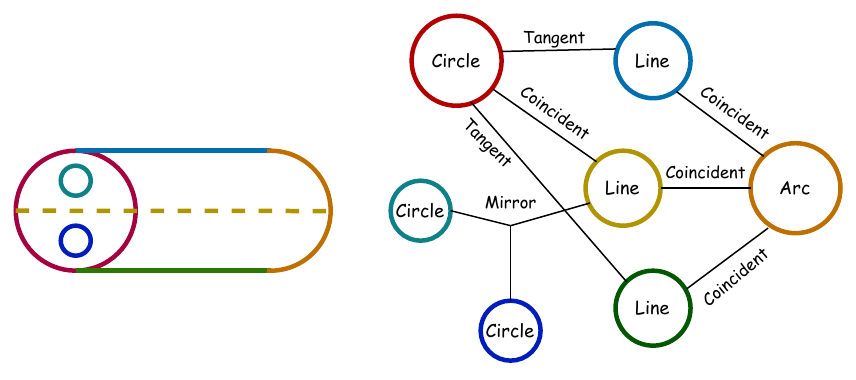}
    \caption{An example of a simple 2D sketch represented as a graph, where the sketch primitives such as curves (circle, arc, line, etc.) are modeled as graph nodes, and the constraints between these primitives are shown as graph edges. SketchGraphs \cite{seff2020sketchgraphs} dataset contains such 2D sketches modeled as graphs. }
    \label{fig:sketchgraph}
\end{figure}

Table \ref{table:datasets} provides a summary of the existing datasets along with their properties. The details about each dataset are provided in the dedicated sections for methodologies linked to each dataset.
Table \ref{table:CADFormats} introduces different CAD data formats along with their corresponding description.
Existing commonly used 3D CAD datasets mostly provide mesh geometry for 3D shape segmentation, classification and retrieval \cite{shilane2004princeton, chang2015shapenet, zhou2016thingi10k, wu20153d, mo2019partnet}, human body mesh registration \cite{bogo2017dynamic}, 2D/3D shape alignment \cite{aubry2014seeing} and 3D scene categorization, semantic segmentation and object detection \cite{song2015sun, dai2017scannet}. 
Engineering shape datasets such as ESB \cite{jayanti2006developing}, MCB \cite{kim2020large}, AAD \cite{bespalov2005benchmarking}, FeatureNet \cite{zhang2018featurenet}, and CADNet \cite{cadnet2021} also provide annotated data with mesh geometry for mechanical shape classification and segmentation. 
The primary limitation of these datasets is their lack of parametric and topological features of curves and surfaces commonly referred to as boundary representation (B-Rep). These B-Rep features are essential for conducting parametric CAD shape analysis. 
Recently, several geometric B-Rep datasets of different sizes and properties were introduced to boost GDL progress on CAD design. FabWave \cite{angrish2019fabsearch} is a collection of $5,373$ 3D shapes annotated with $52$ mechanical part classes, including gears and brackets, and Traceparts \cite{mandelli2022cad} is a small collection of $600$ CAD models produced by different companies labeled into $6$ classes (100 CAD models in each class), including screws, nuts, and hinges, which can be used for 3D shape classification.
MFCAD \cite{cao2020graph} is a synthetic 3D segmentation dataset containing $15,488$ shapes with annotated faces of 16 classes, which can be used for parametric face segmentation in CAD shapes.

Advancing across various tasks in GDL, and effective training of deep learning models, necessitate large parametric CAD data collections. The existing parametric CAD datasets with B-Rep data are limited in size and insufficient to meet these demands. To address this shortfall, three big datasets, ABC \cite{koch2019abc}, Fusion 360 Gallery \cite{willis2021fusion}, and SketchGraphs \cite{seff2020sketchgraphs}, were introduced recently and provide valuable resources for research in this area.
ABC \cite{koch2019abc} is the first large-scale, real-world, and hand designed dataset with over $1$ million high-quality 3D CAD models covering a wide range of object types. Each CAD model consists of a set of precisely parameterized curves and surfaces, offering accurate reference points for sharp geometric feature representation, patch segmentation, analytic differential measurements, and the process of shape reconstruction. The CAD models in ABC dataset are compiled and collected via an openly accessible interface hosted by Onshape \cite{onshape}, and an open-source geometry processing pipeline has been developed to process and prepare the CAD models in ABC dataset to be used by deep learning methods.\footnote{https://deep-geometry.github.io/abc-dataset/}

Training machine learning models to facilitate CAD construction and synthesis can significantly benefit designers by minimizing the time and effort required for the design process.
An essential necessity for tasks related to CAD (re-)construction is to understand how a CAD model is designed and how to interpret the provided construction information in the dataset. The construction history information in ABC dataset can only be retrieved by querying the Onshape API in a proprietary format with limited documentation which makes it challenging to develop CAD (re-)construction methods on that dataset. 
SketchGraphs dataset has been produced to fill this gap by providing a collection of $15$ million human-designed 2D sketches of real-world CAD models from Onshape API. 2D sketches representing geometric primitives (like lines and arcs) and constraints between them (like coincidence and tangency) can be seen as the basis of 3D shape parametric construction, and each 2D sketch is presented as a geometric constraint graph where nodes denote 2D primitives and edges are geometric constraints between nodes (\ref{fig:sketchgraph}).
The SketchGraphs dataset can be used for various applications in automating design process such as auto-completing sketch drawing by predicting sequences of sketch construction operations or interactively suggesting next steps to designer, and autoconstraint application where the method is predicting a set of constraints between geometric primitives of sketch. 
Other potential applications are CAD inference from images where the method receives a noisy 2D drawing image and infers its design steps to produce the corresponding parametric CAD model, and learning semantic representation encoded in sketches.  
An open-source Python pipeline for data processing and preparation for deep learning frameworks also comes along the dataset.\footnote{https://github.com/PrincetonLIPS/SketchGraphs} 
Similar to SketchGraphs, freehand 2D sketch datasets such as \cite{eitz2012humans, sangkloy2016sketchy, gryaditskaya2019opensketch} also have been introduced to tackle this challenge by providing the sketch construction sequence. 

Fusion 360 Gallery \cite{willis2021fusion} has been introduced recently by Autodesk as the first human designed 3D CAD dataset and environment with 3D operation sequences for programmatic CAD construction. The fusion 360 Gallery dataset comes along an open source Python environment named as Fusion 360 Gym for processing and preparing the CAD operations for machine learning methods. 
This dataset contains both 2D and 3D geometric CAD data which are produced by the users of the Autodesk Fusion 360 CAD software and are collected into the Autodesk Online Gallery. Several datasets for different learning goals, such as CAD reconstruction \cite{willis2021fusion}, CAD segmentation \cite{lambourne2021brepnet}, and CAD assembly \cite{willis2022joinable}, are created based on a total of real-world $20,000$ designs in Fusion 360 Gallery. 

\begin{table*}[ht]
	\centering
	\resizebox{\linewidth}{!}{
	\begin{tabular}{lcccccc}
        \hline
		Dataset  & \#Models & B-Rep & Mesh & Sketch & \#Categories & Application  \\ 
		\hline
        ShapeNet \cite{chang2015shapenet} & 3M+ & $\times$ & \checkmark & $\times$ & 3,135 & Classification  \\ 
        ModelNet \cite{wu20153d} & 12,311 & $\times$ & \checkmark & $\times$ & 40 & Classification  \\ 
        PartNet \cite{mo2019partnet} & 26,671 & $\times$ & \checkmark & $\times$ & 24 & Classification, Segmentation  \\ 
        PrincetonSB \cite{shilane2004princeton} & 6,670 & $\times$ & \checkmark & $\times$ & 92 & Classification  \\ 
        \hline \hline
        AAD \cite{bespalov2005benchmarking} & 180 & $\times$ & \checkmark & $\times$ & 9 & Classification \\ 
        ESB \cite{jayanti2006developing} & 867 & $\times$ & \checkmark & $\times$ & 45 & Classification  \\ 
        Thingi10k \cite{zhou2016thingi10k} & 10,000 & $\times$ & \checkmark & $\times$ & 2,011 & Classification \\ 
        FeatureNet \cite{zhang2018featurenet} & 23,995 & $\times$ & \checkmark & $\times$ & 24 & Classification \\ 
        MCB \cite{kim2020large} & 58,696 & $\times$ & \checkmark & $\times$ & 68 & Classification  \\ 
        CADNet \cite{cadnet2021} & 3,317 & $\times$ & \checkmark & $\times$ & 43 & Classification  \\ 
        
        FABWave \cite{angrish2019fabsearch} & 5,373 & \checkmark & \checkmark & $\times$ & 52 & Classification  \\ 
        Traceparts \cite{mandelli2022cad} & 600 & \checkmark & \checkmark & $\times$ & 6 & Classification \\ 
        SolidLetters \cite{jayaraman2021uv} & 96,000 & \checkmark & $\times$ & $\times$ & 26 & Classification\\ 

        \hline \hline
        MFCAD \cite{cao2020graph} & 15,488 & \checkmark & $\times$ & $\times$  & 16 & Segmentation \\ 
        MFCAD++ \cite{colligan2022hierarchical} & 59,655 & \checkmark & $\times$ & $\times$  & 25 & Segmentation \\
        Fusion 360 Segmentation \cite{lambourne2021brepnet} & 35,680 & \checkmark & \checkmark & $\times$  & 8 & Segmentation \\ 
        CC3D-Ops & 37,000 & \checkmark & \checkmark & $\times$ & - & Segmentation\\

        \hline \hline
        Fusion 360 Assembly \cite{willis2022joinable} & 154,468 & \checkmark & \checkmark & $\times$ & $-$ & Joint Prediction\\  
        AutoMate \cite{jones2021automate} & 3M+ & \checkmark & \checkmark & $\times$ & $-$ & Joint Prediction\\ 
        
        \hline \hline
        ABC \cite{koch2019abc} & 1M+ & \checkmark & \checkmark & $\times$ & $-$ & Reconstruction  \\ 
        Fusion 360 Reconstruction \cite{willis2021fusion} & 8,625 & \checkmark & \checkmark & $\times$ & $-$ & Reconstruction\\ 
        SketchGraphs \cite{seff2020sketchgraphs} & 15M+ & $\times$ & $\times$ & \checkmark & $-$ &  Reconstruction  \\ 
        CAD as Language \cite{ganin2021computer} & 4.7M+ & $\times$ & $\times$ & \checkmark & $-$ & Reconstruction  \\ 
        Sketch2CAD \cite{li2020sketch2cad}  & 50,000 & $\times$ & $\times$ & \checkmark  & $-$ & Reconstruction\\
        CAD2Sketch \cite{hahnlein2022cad2sketch}  & 6000 & $\times$ & $\times$ & \checkmark & $-$ & Reconstruction\\
        DeepCAD \cite{wu2021deepcad} & 178,238 & \checkmark & $\times$ & $\times$ & $-$ & Reconstruction\\
        PVar \cite{solidgen23} & 120,000 & \checkmark & $\times$ & $\times$ & 60 & Reconstruction, Classification\\
        CADParser \cite{zhou2023cadparser}  & 40,000 & \checkmark & $\times$ & $\times$ & $-$ & Reconstruction\\
        Free2CAD \cite{li2022free2cad} & 82,000 & \checkmark & $\times$ & \checkmark & $-$ & Reconstruction\\
        \hline 
    \end{tabular}
    }
\caption{Overview of the existing common object and mechanical CAD datasets with their properties. For each dataset, the number of CAD models, representation formats such as \textit{B-Reps, Mesh, Sketch}, and different tasks they are annotated for, such as Segmentation, Classification and CAD Reconstruction, are reported. The first 4 rows of the table show the common object datasets annotated for 3D shape classification. The remaining rows list mechanical object datasets, created for CAD analysis.} 
\label{table:datasets}
\end{table*}

\section{CAD Representation Learning} 
\label{sec:representation_learning}

Studying extensive data repositories and uncovering hidden features in data for various tasks such as similarity analysis and shape retrieval has been a vibrant field of research in machine learning and artificial intelligence. The significance of this concept extends to CAD data as well. Machine learning-based similarity analysis of CAD models can effectively facilitate the design process for designers by categorizing designs, and retrieving similar CAD models as design alternatives. It has been shown that approximately 40\% of new CAD designs could be constructed based on existing designs in the CAD repository, and at least 75\% of design processes leverage existing knowledge for designing a new CAD model \cite{gunn1982mechanization}. Learning from CAD data and extracting features from extensive collections of geometric CAD shapes is important for CAD retrieval and similarity analysis, and it involves employing various machine learning and deep learning methods for extracting geometric attributes encoded in B-Rep data for assessing similarity between pairs of CAD models. 

The first step in this process is to select a subset of geometrical, topological, functional and other properties of the CAD model, based on specific analysis goals, to be represented in a suitable format, such as numerical vectors, matrices, tensors, or graphs, for the deep learning methods to process. 
Nevertheless, representing B-Rep data is challenging due to the coexistence of continuous non-Euclidean geometric features and discrete topological properties, making it difficult to fit into regular structured formats such as tensors or fixed-length encodings.
A key contribution of each state-of-the-art method in this topic is the introduction of a method for encoding or tokenizing B-Rep data into a format suitable for the adopted deep learning architecture, tailored to the particular application they are addressing. 
The deep learning methods receive these representations as input and learn to classify, cluster, segment, or reconstruct CAD models, considering the annotations at hand. However, the application of machine learning and deep learning methods to CAD models is very rare because of the scarcity of annotated CAD data available in the B-Rep format. In contrast to the geometric CAD models like ShapeNet \cite{chang2015shapenet}, many parametric CAD datasets are not publicly released due to the proprietary nature of design data. 

While there have been recent releases of small annotated datasets containing mechanical parts in B-Rep format for machine learning research \cite{mandelli2022cad}, the majority of large-scale public databases, such as ABC, remain predominantly unlabeled. Additionally, not only the process of manually annotating B-Rep data in a specialized format needs engineering expertise, but it is also very time-consuming and costly, thus posing a significant limitation. 
Consequently, deep learning approaches that do not rely on external annotations, such as unsupervised and self-supervised learning, become particularly crucial alongside traditional supervised learning approaches in such scenarios. Learning feature representations based on intrinsic data features, without external annotations or expert knowledge, proves highly advantageous in overcoming the scarcity of annotated CAD data.
In this section, we introduce existing research endeavors that employ GDL to extract feature representations from CAD B-Reps. These methods operate in either a supervised, self-supervised, or unsupervised manner, catering to various tasks such as classification, segmentation, similarity analysis and retrieval, and assembly prediction.

\subsection{CAD Classification and Retrieval}
Designing methodologies for categorizing 3D components within B-Rep assemblies is significantly important for various applications, such as re-using similar CAD components in different assemblies, shape recommendation and alternative suggestion, and product complexity estimation, especially when 3D CAD models with substantially different geometries are within the same category and share similar topology. 
One of the first deep learning methods proposed to work directly on B-Rep data format of 3D CAD models is UV-Net \cite{jayaraman2021uv}. UV-Net proposed a cohesive graph representation for B-Rep data by modeling topology with an adjacency graph, and modeling geometry in a regular grid format based on the U and V parameter domains of curves and surfaces. 
One of the main contributions of this work is to extract crucial geometric and topological features out of B-Rep data and make a grid data structure out of complex B-Rep data to feed to deep learning models for different tasks such as B-Rep classification and retrieval. 
To generate grid-structured feature representations from B-Rep data, this approach transforms each 3D surface into a regular 2D grid by performing surface sampling with fixed step sizes. In a similar fashion, it transforms each solid curve into a 1D grid. The resulting 1D/2D grid mappings are referred to as UV-grids. Each sampled point in a surface's 2D grid conveys three distinct values across $7$ channels: a) The 3D absolute point position, represented as $xyz$ in the UV coordinate system, b) The 3D absolute surface normal, and c) The trimming mask with $1$ and $0$ denoting samples in the visible region and trimmed region, respectively. For the sampled points in a curve's 1D grid, the encoding includes the absolute point UV coordinates and, optionally, the unit tangent vector.

\begin{figure*}[ht]
    \centering
    \includegraphics[width=\textwidth]{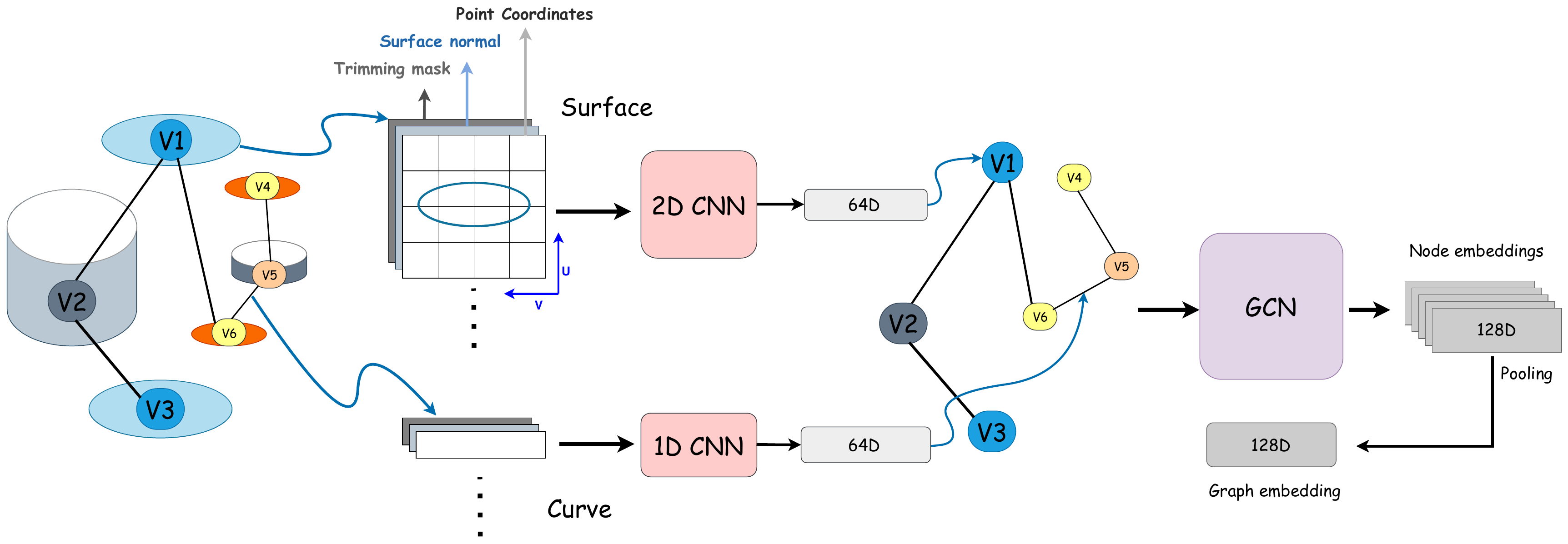}
    \caption{Schematic representation of the UV-Net model architecture \cite{jayaraman2021uv}. The model takes B-Rep data as input, generating grid-structured features for surfaces and connecting curves. These UV-grid mappings are further processed through CNN and GCN architectures to learn feature embeddings for the overall graph and its individual nodes.}
    \label{fig:uvnet}
\end{figure*}

As illustrated in Figure \ref{fig:uvnet}, UV-Net model architecture is comprised of both CNN and GCN layers to first extract features from 1D and 2D grids representing curves and surfaces, respectively, and then to capture the topological structure of the 3D shape encoded as a graph with hierarchical graph convolution layers. 
The 1D curve and 2D surface UV-grids are processed by 1D and 2D convolution and pooling layers while the weights of the convolution layers are shared among all curves and surfaces in a B-Rep to make them permutation-invariant. 
The $64$-dimensional feature vectors derived from surfaces and curves by the convolution layers of the CNN serve as the node and edge features within a face-adjacency graph $\mathcal{G}(\mathcal{V}, \mathcal{E})$, where the set of graph nodes $\mathcal{V}$ represent the faces (surfaces) in B-Rep and the set of edges $\mathcal{E}$ represents the connection between faces. 
Subsequently, this graph is introduced to a multi-layer GCN, where the graph convolution layers propagate these features across the graph, enabling the capturing of both local and global structures inherent in the shape. 
The UV-Net encoder is used as a backbone for supervised and self-supervised learning on labeled and unlabeled datasets, respectively. 
For CAD classification, which is a supervised task, UV-Net encoder is followed by a 2-layer classification network to map the learned features to classes and the model is trained in an end-to-end manner on $3$ annotated datasets, SolidLetters \cite{jayaraman2021uv}, FabWave \cite{angrish2019fabsearch}, and FeatureNet \cite{zhang2018featurenet}. 
SolidLetters is currently the biggest synthetic 3D B-Rep dataset with a great variation in both geometry and topology annotated for a classification task. 
It consists of $96,000$ 3D shapes which represent $26$ English alphabet letters (a-z) with different fonts and dimensions.

For CAD retrieval, however, the premise is the absence of labeled data, and thus, the UV-Net encoder needs to be trained in a self-supervised manner. This leads to the utilization of deep learning models designed for self-supervised training, such as Graph Contrastive Learning (GCL) \cite{you2020graph} or Graph AutoEncoder (GAE) \cite{li2018learning}. 
For training the UV-Net encoder as a self-supervised model, GCL is leveraged to apply transformations on the face-adjacency graph and make positive pairs for each B-Rep sample. These transformations can be performed in various ways, such as randomly deleting nodes or edges with uniform probability or extracting a random node and its n-hop neighbors within a graph. 
Assuming that each B-Rep and its transformed verison are positive pairs, the UV-Net encoder extracts the shape embeddings of each pair as $\{h_i, h_j\}$ and a 3-layer non-linear projection head with ReLU activations transforms these embeddings into latent vectors $z_i$ and $z_j$, respectively. 
For a batch comprising $N$ B-Rep samples, the latent vector for each sample, along with its corresponding positive pair, is computed. The entire model is then trained in an end-to-end manner with the objective of bringing the embedding of each sample closer to its positive pair. Simultaneously, the model works to treat the remaining $2(N-1)$ B-Reps as negative examples, aiming to push them further away from the positive pair.
Through this process, the model learns to capture the intrinsic features and patterns within the data without the need for labeled examples. It essentially leverages the relationships within the data itself, generated through transformations or augmentations, to learn meaningful representations. This makes contrastive learning a self-supervised learning approach which is particularly useful in scenarios where obtaining labeled data is challenging or impractical. For retrieving similar CAD models, the model is trained on an unlabeled dataset like ABC. Subsequently, embeddings for random samples from the test set are used as queries, and their k-nearest neighbors are calculated within the UV-Net shape embedding space.

Likewise, the method proposed in \cite{jones2023self} utilizes geometry as a means of self-supervision, extending its application to few-shot learning.
Specifically, the method involves training an encoder-decoder structure to rasterize local CAD geometry, taking CAD B-Reps as input and generating surface rasterizations as output.
B-Reps are assembled piecewise through explicitly defined surfaces with implicitly defined boundaries. Accordingly, the encoder of this approach adopts the hierarchical message passing architecture of SB-GCN proposed in \cite{jones2021automate} to effectively capture boundary features for encoding B-Rep faces. The decoder, in turn, reconstructs faces by simultaneously decoding the explicit surface parameterization and the implicit surface boundary.
The embeddings learned through self-supervised learning in this method serve as input features for subsequent supervised learning tasks, including CAD classification on the FabWave dataset. Notably, with very limited labeled data (ranging from tens to hundreds of examples), the method outperforms previous supervised approaches while leveraging smaller training sets. This underscores the effectiveness of the differentiable CAD rasterizer in learning intricate 3D geometric features.

\subsection{CAD Segmentation}

\begin{figure}
    \centering
    \includegraphics[width=1\linewidth]{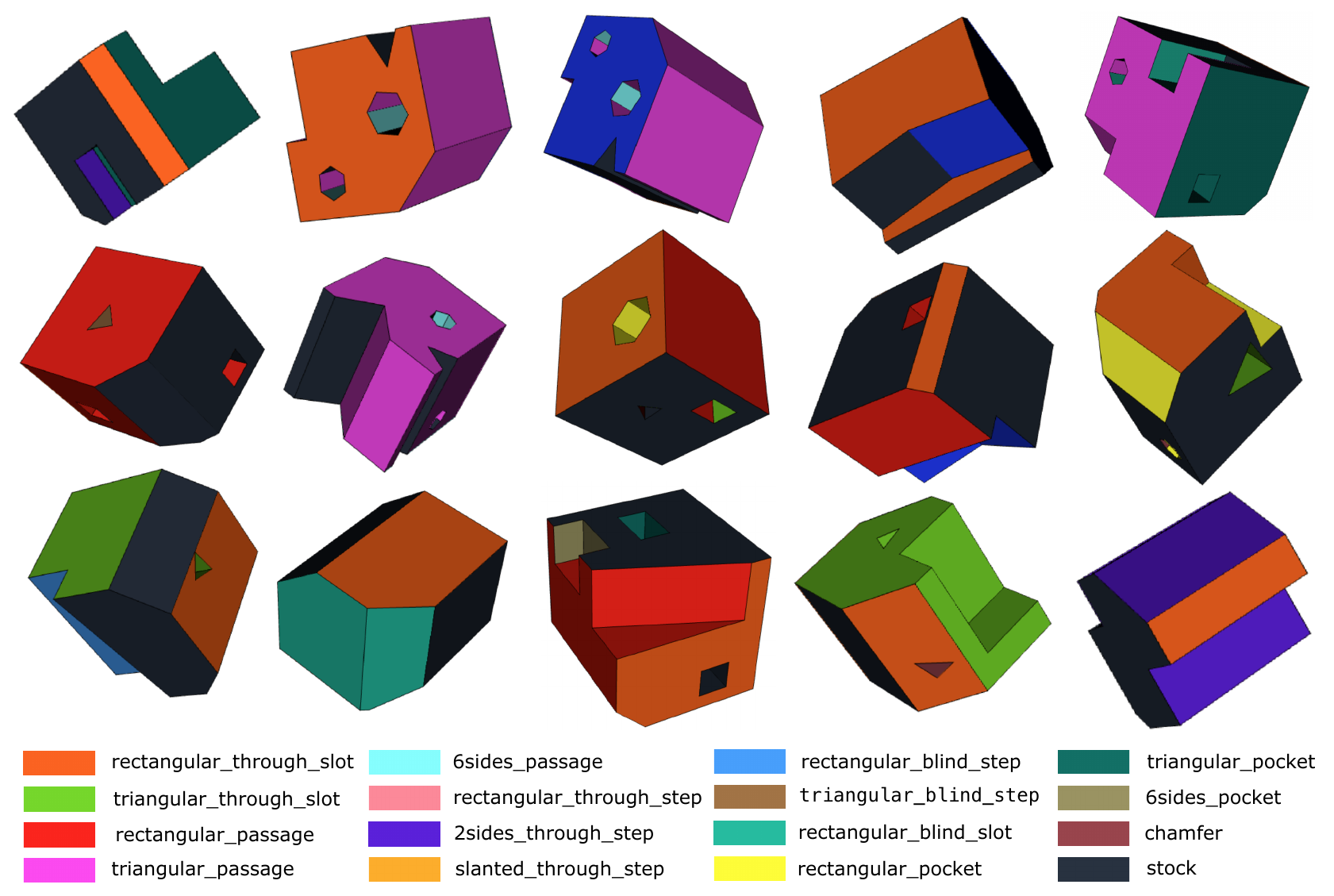}
    \caption{Examples from the MFCAD dataset for manufacturing driven segmentation. Each solid face is labeled with $16$ types of machining features \cite{cao2020graph}.}
    \label{fig:seg_mfcad}
\end{figure}

CAD segmentation entails decomposing a geometric solid model, represented in B-Rep format, into its individual components including faces and edges. 
CAD segmentation finds applications in retrieving CAD parametric feature history, machining feature recognition, Computer Aided Engineering (CAE) and Computer Aided Process Planning (CAPP). 
This is particularly intriguing as it facilitates the automation of several tedious manual tasks within CAD design and analysis, especially when users need to repeatedly select groups of faces and/or edges according to the manufacturing process as input for modeling or manufacturing operations \cite{al2018survey, shi2020critical, cao2020graph}.
However, progress in CAD segmentation was hindered until very recently due to the absence of advanced deep learning methods and large annotated datasets.
MFCAD \cite{cao2020graph} is the first synthetic segmentation dataset consisting of $15,488$ 3D CAD models with planar faces, each annotated with $16$ types of machining feature such as chamfer, triangular pocket, through hole, etc. Figure \ref{fig:seg_mfcad} shows some examples from this dataset. 

The CAD segmentation task can be framed as a node classification problem, with each 3D solid represented as a face adjacency graph. The graph nodes, corresponding to B-Rep faces, are then classified into distinct machining feature classes. 
CADNet \cite{colligan2022hierarchical} is one of the first proposed methods in this regard which represents the B-Rep solid as a graph encoding face geometry and topology, and utilizes a hierarchical graph convolutional network called Hierarchical CADNet to classify the graph nodes (or solid faces) into different machining features. For evaluation, this method not only leveraged MFCAD dataset, but it also created and released the extended version of the dataset, MFCAD++, which consists of $59,655$ CAD models with $3$ to $10$ machining features including both planar and non-planar faces.
The UV-Net method, as introduced in the previous section, also addresses CAD segmentation in the same fashion, training its encoder in a supervised manner on the MFCAD dataset.
However, translating the B-Rep into a face adjacency graph results in the loss of some information regarding the relative topological locations of nearby entities.
Moreover, graph representations constructed based on UV coordinates lack invariance to translation and rotation.

\begin{figure}
    \centering
    \includegraphics[width=0.7\linewidth]{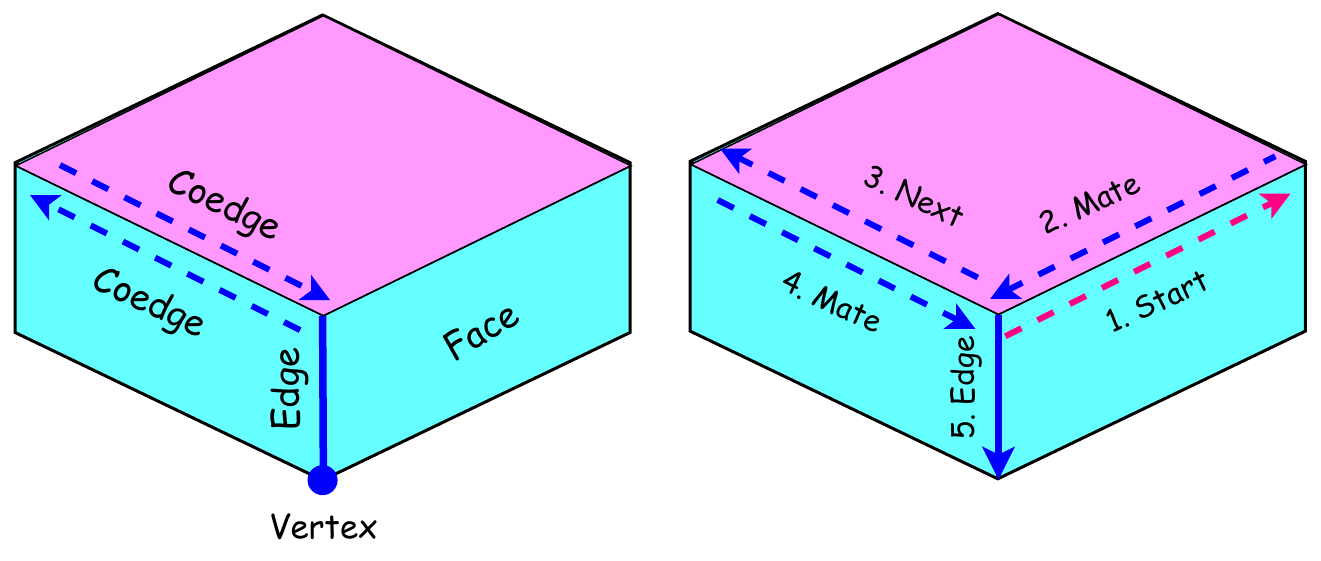}
    \caption{(left): The topology of a solid can be defined as a set of face, edge, coedge, vertex entities. and it can be traveresed by a sequence of walks. (right): The topology can be traveresed by a sequence of walks from the starting entitiy (in red). The walks can be \textit{Edge}, \textit{Face}, \textit{Next}, \textit{Previous} and \textit{Mate}. As an example, here the starting entity is the red coedge, followed by \textit{Mate}, \textit{Next}, \textit{Mate}, \textit{Edge} walks \cite{lambourne2021brepnet}.}
    \label{fig:coedge}
\end{figure}

\begin{figure}
    \centering
    \includegraphics[width=1\linewidth]{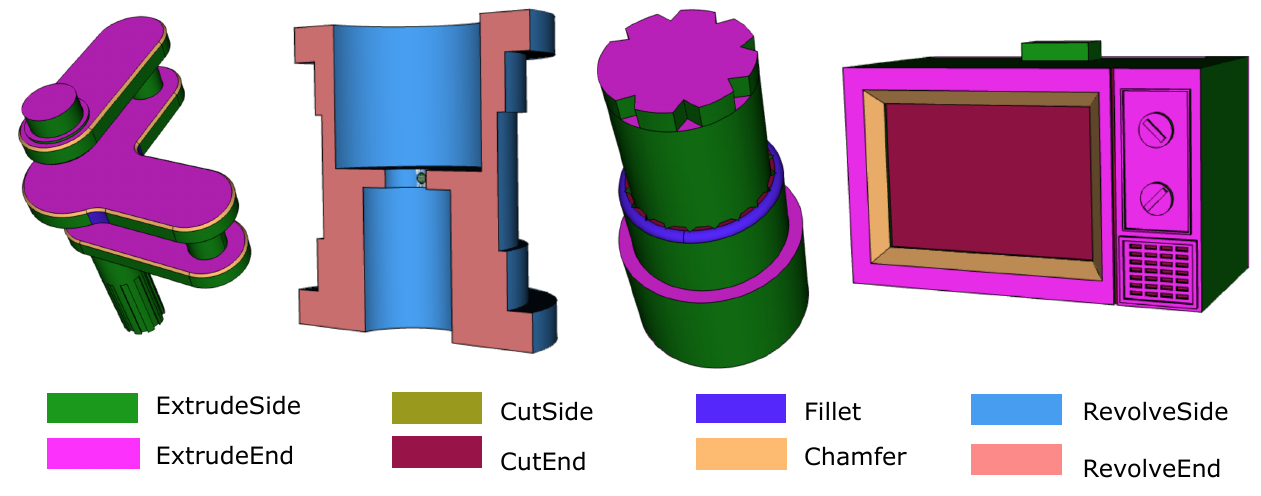}
    \caption{Examples from the Fusion 360 Gallery dataset annotated for construction-based segmentation. Each solid face is labeled with operation used in its construction \cite{lambourne2021brepnet}.}
    \label{fig:seg_fusion360}
\end{figure}

BRepNet \cite{lambourne2021brepnet} is the first method specifically designed for B-Rep segmentation based on deep learning and, notably, it accomplishes this without introducing to the network any coordinate information. It operates directly on B-Rep faces and edges, exploiting compact information derived from their topological relationships for B-Rep segmentation.
The motivation behind the BRepNet approach stems from the idea seen in the convolutional operation in CNNs used for image processing. In this operation, the local features are aggregated by sliding a small window called a filter or kernel over the grid data and performing element-wise multiplication between the filter and the overlapping grid cells, then pooling the results.
This concept is extended to B-Reps, allowing the identification of a collection of faces, edges, and coedges at precisely defined locations relative to each coedge in the data structure.
Coedge is a doubly linked list of directed edges, representing the neighboring structures of B-Rep entities. Each coedge also retains information about its parent face and edge, its neighbor (mating) coedge, and also pointers to the next and previous coedge in the loop around a face. 
Figure \ref{fig:coedge} illustrates a topology example traversed by a sequence of walks from a given coedge (in red) to some of its neighboring entities such as mating coedge, next and previous coedges, faces and edges. 
Information about geometric features around a coedge, including face and edge type, face area, edge convexity and length, and coedge direction, is encoded as one-hot vectors and concatenated in a predetermined order forming feature matrices $X^f$, $X^e$, $X^c$ of face, edge and coedge, respectively. These matrices are then passed into a neural network where convolution operations are performed through matrix/vector multiplication to recognize patterns around each coedge.
Additionally, the performance of BRepNet is evaluated on a segmentation dataset from the Fusion 360 Gallery which is released alongside BRepNet method, as the first segmentation dataset comprising of real 3D designs. As shown in Table \ref{table:datasets}, this dataset comprises $35,680$ 3D shapes each annotated with $8$ types of modeling operations utilized to create faces in the respective model. Figure \ref{fig:seg_fusion360} shows some examples from this dataset. 

The self-supervised method proposed in \cite{jones2023self}, is also assessed on a segmentation task by initially pre-training the network on a subset of $27,450$ parts from the Fusion 360 Gallery segmentation dataset. This pre-training is conducted in a self-supervised manner without utilizing face annotations. Subsequently, the pre-trained network undergoes fine-tuning in a supervised manner, being exposed to only a limited number of annotated parts to demonstrate the method's performance in a few-shot setting on both Fusion 360 and MFCAD datasets.
CADOps-Net \cite{dupont2022cadops} draws inspiration from CAD segmentation methods like UV-Net \cite{jayaraman2021uv} and BRepNet \cite{lambourne2021brepnet}, which segment B-Reps into distinct faces based on their associated CAD operations, and proposes a neural network architecture that takes the B-Rep data of a 3D shape as input and learns to decompose it into various operation steps and their corresponding types. Besides, it introduces the CC3D-Ops dataset, comprising $37,000$ CAD models annotated with per-face CAD operation types and construction steps \footnote{https://cvi2.uni.lu/cc3d-ops/}.

\subsection{CAD Assembly}

The physical objects we encounter all around us are primarily intricate assemblies built up by CAD designers by designing and then aligning multiple smaller and simpler parts through CAD software. The meticulous pairing of parts in CAD is a laborious manual process, consuming approximately one-third of designers' time \cite{jones2021automate}. It entails precise positioning of parts in relation to each other and specifying their relative movement. Consequently, optimizing this process is crucial for enhancing the efficiency of CAD systems. 
This issue has been explored in various studies using deep learning methods in recent years \cite{zhan2020generative, lin2018learning, mo2019structurenet, jones2020shapeassembly, wu2019sagnet, yin2020coalesce, zhu2018scores, harish2022rgl} to simplify the part assembling and open avenues for various applications such as robotic assembly \cite{lee2021ikea}, CAD assembly synthesis \cite{sung2017complementme}, part motion prediction \cite{wang2019shape2motion}, robot design optimization \cite{zhao2020robogrammar} and similarity analysis in CAD assemblies \cite{boussuge2019application}. 
However, all these approaches operate with non-parametric data structures like meshes, point clouds, and voxel grids. Consequently, they leverage GDL methods tailored for these data structures, such as DGCNN \cite{wang2019dynamic}, PCPNet \cite{guerrero2018pcpnet}, PointNet \cite{qi2017pointnet}, and PointNet++ \cite{qi2017pointnet}, to learn the surface representations. These methods primarily adopt a top-down approach to predict the absolute pose of a set of parts for assembly in a global coordinate system. However, this approach lacks support for the parametric variations of parts for modifying the assembly or modeling degrees of freedom, and it may also result in failures when parts cannot achieve complete alignment. Additionally, these methods heavily rely on annotated datasets like PartNet \cite{mo2019partnet}, which exclusively provides data in mesh format.

A bottom-up approach to assembling pairs of parts, relying on pairwise constraints and utilizing the joint or contact information available in B-Rep data, can address this issue without requiring class annotations on datasets. However, current B-Rep datasets like ABC \cite{koch2019abc} and Fusion 360 Gallery \cite{willis2021fusion} lack pairing benchmarks for CAD assemblies, making them unsuitable for training assembly prediction models.
AutoMate \cite{jones2021automate} and JoinABLe \cite{willis2022joinable} are the only two works proposed recently, employing a bottom-up learning approach for pairing parts locally to form assemblies. Additionally, they have released B-Rep datasets along with their methods, providing pairing benchmarks for training models.

\subsubsection{AutoMate \cite{jones2021automate}}
Automate is the first work in this area focusing on CAD assembly by operating on parametric B-Rep data format. Additionally, this work introduces the first large-scale dataset of CAD assemblies in B-Rep format with mating benchmarks, released to facilitate future research in this field. AutoMate dataset is made by collecting publicly available CAD designs from Onshape API, containing $92,529$ unique assemblies with an average size of $12$ mates each, and $541,635$ unique mates. Mating in this work means aligning two parts with pairwise constraints defined based on B-Rep topology. 
These pairwise constraints are referred to as mates or joints, and they dictate the relative pose and degrees of freedom (DOF) of parts within an assembly.
Two parts can be mated through various topological entities, such as faces, edges, and vertices. Therefore, it is essential to learn the feature representation of multiple topological entities at different levels to address the complexities of the CAD assembly problem.
In contrast to previous CAD representation learning approaches, like BRepNet \cite{lambourne2021brepnet} and UV-Net \cite{jayaraman2021uv}, which construct a face adjacency graph to capture the homogeneous structure of the B-Rep and focus on learning feature representations for face entities, AutoMate \cite{jones2021automate} takes a different approach. It aims to learn representations for faces, loops, edges, and vertices by capturing the heterogeneous B-Rep structure. This is achieved through the introduction of the Structured B-Rep Graph Convolution Network (SB-GCN) architecture, a message-passing network designed for learning the heterogeneous graph representation of the B-Rep. 

SB-GCN takes a heterogeneous graph as input, where faces $\textit{F}$, edges $\textit{E}$, vertices $\textit{V}$, and loops $\textit{L}$ serve as graph nodes, and directed bipartite connection sets between them represent graph edges. Specifically, the relations between B-Rep vertices and edges are denoted by $\textit{V:E}$ and its transpose $\textit{E:V}$. Similarly, $\textit{E:L}$ and $\textit{L:E}$ represent relationships between B-Rep edges and loops, while $\textit{L:F}$ and $\textit{F:L}$ denote connections between faces and loops. Additionally, there are undirected relations (meta-path) between geometrically adjacent faces, expressed as $\textit{F:F}$. Each node is associated with a parametric geometry function encoded as a one-hot feature vector.
The network utilizes structured convolutions to generate output feature vectors for all graph nodes. The adjacency structure of different node types is captured in an ordered hierarchy in different network layers. The initial three layers capture relations in the B-Rep hierarchy in a bottom-up order: Vertex to Edge, Edge to Loop, and Loop to Face. Subsequently, the following $k$ layers focus on capturing meta-relations between faces. The final three layers reverse the node relations: Face to Loop, Loop to Edge, and Edge to Vertex.
The network makes predictions for mate location and type using two distinct output heads. The mate location prediction head assesses pairs of mating coordinate frames (MCFs) that are adjacent to the selected faces on the two mating parts. Meanwhile, the mate type prediction head anticipates how this pair of MCFs should be mated. This is done by classifying features into seven different mating categories, i.e. \textit{fastened},\textit{revolute}, \textit{planar}, \textit{slider}, \textit{cylindrical}, \textit{parallel}, \textit{ball}, \textit{pin slot}.
The model is trained on $180,102$ mates from the AutoMate dataset, and is integrated as an extension to the Onshape CAD system. This integration assists designers by providing mate recommendations for pairs of parts during the CAD design process.

\subsubsection{JoinABLe \cite{willis2022joinable}}
Joinable is another recent method that employs a bottom-up approach to predict joints (or mates) between parts based on pairwise constraints. In contrast to AutoMate, which relies on user contact surface selections on parts to rank and recommend multiple mating solutions, JoinABLe identifies joints between parts without being limited to predefined surfaces, doing so automatically and without requiring any user assistance.
The Fusion 360 Gallery Assembly dataset is concurrently introduced and made available with the JoinABLe method for the purpose of training and evaluating the JoinABLe model. This dataset comprises two interconnected sets of data: Assembly and Joint. These datasets are collected from user-designed CAD models publicly accessible in the Autodesk Online Gallery. The Assembly data encompasses $8,251$ assemblies, totaling $154,468$ individual parts, along with their corresponding contact surfaces, holes, joints, and the associated graph structure. The Joint data includes $23,029$ distinct parts, incorporating $32,148$ joints between them. JoinABLe is trained using the joint information extracted from Fusion 360 Gallery Assembly-Joint dataset. 
The Assembly Dataset \cite{willis2022joinable} is a subset of designs in which each CAD model is made of multiple parts. In CAD design context, each assembly or 3D shape is a collection of parts that joined together and a set of assemblies can make a CAD design or 3D object.

\begin{figure}
    \centering
    \includegraphics[width=0.6\linewidth]{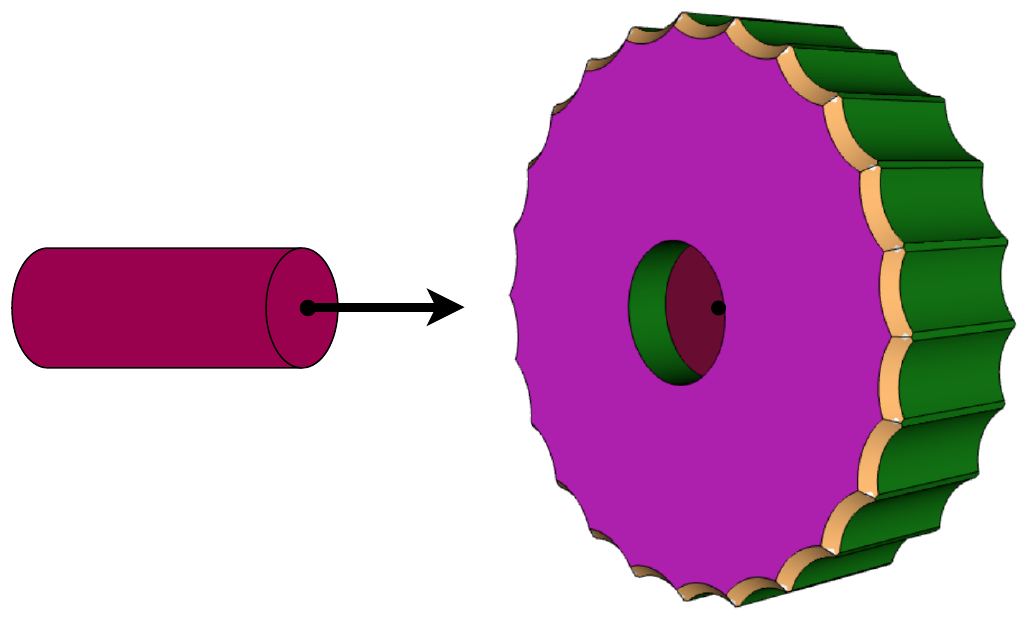}
    \caption{An example showing how joints are defined between parts (solids) in Fusion 360 Assembly-Joint dataset. Give a pair of parts, JoinABLe predicts the joint axis and pose between them according to the joint defined by the ground truth \cite{willis2022joinable}.}
    \label{fig:fusion360_assembly}
\end{figure}

Two parts are aligned through their joint axes, each possessing a specific direction. The joint axis on each part can be defined on either a face or edge entity, featuring an origin point and a directional vector. For instance, on a circular surface entity, the origin point is the center of the circle and the direction vector is the normal vector. These joint axes information for pairs of parts are provided in Fusion 360 Gallery Assembly-Joint dataset, which is a subset of Fusion 360 Gallery Assembly dataset, as ground truth for training a joint prediction deep learning model. This dataset contains 23,029 parts with $32,148$ joints between them. 
As shown in Figure \ref{fig:fusion360_assembly}, given two parts, JoinABLe is designed to predict the parametric joint between them, encompassing joint axis prediction (origin points on two parts and direction vectors) and joint pose prediction. To facilitate this, each part is modeled as a graph $\mathcal{G}(\mathcal{V}, \mathcal{E})$ with B-Rep faces and edges serving as graph nodes, and their adjacency relationships functioning as graph edges.
B-Rep faces are encoded as one-hot feature vectors, representing surface types like plane and cylinder, along with a flag indicating if the surface is reversed relative to the face. Similarly, one-hot vectors for B-Rep edges encompass curve characteristics such as line and circle, curve length, and a flag denoting whether the curve is reversed concerning the edge. These one-hot vectors function as graph node features.

For two parts represented by graphs $G_1$ and $G_2$ with $N$ and $M$ nodes, respectively, the joint graph $G_j$ is constructed to illustrate connections between the two graphs. This connection information in $G_j$ can be expressed as a binary matrix of dimensions $N \times M$. In alignment with the ground truth joints provided in the Joint dataset, only one matrix element, representing the joint between two entities across the two parts, should be set to $1$, while the others remain $0$. The JoinABLe model aims to find that one positive element and it comprises three main components: encoder, joint axis prediction, and joint pose prediction.
The encoder follows a Siamese-style architecture with two distinct MLP branches. One branch focuses on learning node features that represent B-Rep faces, while the other concentrates on learning node features representing B-Rep edges. The learned embeddings for faces and edges in each part are subsequently concatenated to form the node embeddings for the corresponding graph. These node embeddings are then input into a Graph Attention Network v2 (GATv2) \cite{brody2021attentive} which captures the local neighborhood structure within each graph through message passing.

The joint axis prediction is formulated as a link prediction task, aiming to predict a connection between the graphs $G_1$ and $G_2$ by linking two nodes.  This involves edge convolution on the joint graph $G_j$ which illustrates the connections between the two graphs and is updated with the node features learned by the encoder in the previous step. Assuming that $h_v$, $h_u$ represent the learned features for the two nodes $v$, $u$, from $G_1$, $G_2$, respectively, the edge convolution to learn the connection between the two nodes in the two graphs is performed as follows: 
\begin{equation}
    h_{uv} = \Phi(h_u \oplus h_v), 
\end{equation}
where $\Phi(\cdot)$ is a 3-layer MLP applied to the concatenated features of $h_u$, $h_v$. Following the edge convolution, a softmax function is applied to normalize the features and predict the most probable link between the nodes in $G_j$. 
After predicting the joint axes to align the two parts, the pose prediction head employs a neurally guided search approach to iterate through the top-k joint axis predictions. 
As an supplementary work, JoinABLe has also proposed assembling a multi-part design using only the individual parts and the sequence of part pairs derived from the Assembly dataset. However, in real-world applications, this well-defined assembly sequence and the corresponding assembly graph might not be available for the network to use. Additionally, a misalignment in any of the assembly steps could result in an incorrect overall assembly. Therefore, it is suggested that for large and complex assemblies, a combination of top-down and bottom-up approaches might be more effective.

In the CAD Assembly problem two primary questions often arise: how to select the pair of parts to joint (or mate), and how to assemble them \cite{funkhouser2004modeling}. 
The former question can be addressed by leveraging similarity analysis and retrieval methods to identify suitable pairs of parts. AutoMate and JoinAble, on the other hand concentrate on answering the latter question by learning the process of mating two parts in an assembly.

\section{CAD Construction with Generative Deep Learning} 

CAD construction with generative deep learning involves leveraging advanced GDL methods to automatically generate or assist in the creation of parametric CAD models. These approaches can support designers in various ways to streamline the design process. This includes tasks like generating or auto-completing sketches, as well as generating CAD operations to construct a 3D model based on the designed sketch.
In this section, we categorize these methods into six distinct groups: 1) methods focusing on 2D sketches, aiming to automate the sketching process in a 2D space as an initial step before developing 3D CAD models, 2) methods targeting 3D CAD model reconstruction given the sketch of the model, 3) methods generating CAD construction sequences, specifically focused on sketch and extrude operations, for CAD construction, 4) methods which perform direct B-Rep synthesis to generate 3D CAD models, 5) methods generating 3D CAD models from Point Cloud data, and 6) methods generating CAD models from images.

\subsection{Engineering 2D Sketch Generation for CAD}
\label{subsec:2dsketchgen}
Engineering 2D sketches form the basis 3D CAD design. The 2D sketches composed of a collection of geometric primitives, such as vertex, lines, arcs, and circles, with their corresponding parameters (e.g. radius, length, and coordinates), and explicit imposed constraints between primitives (e.g. perpendicularity, orthogonality, coincidence, parallelism, symmetry, and equality) determining their final configuration. These 2D sketches can then be extruded to make 3D designs. 
Synthesizing parametric 2D sketches and learning their encoded relational structure can save a lot of time and effort from designers when designing complex engineering sketches. However, leveraging deep learning approaches in this regard needs large-scale datasets of 2D engineering sketches. Most of the existing large-scale datasets provide hand-drawn sketches of common objects, such as furniture, cars, etc. The QuickDraw dataset \cite{ha2017neural} is collected from the Quick, Draw! online game \cite{jongejan2016quick}, and Sketchy dataset \cite{sangkloy2016sketchy} is a collection of paired pixel-based natural images and their corresponding vector sketches. These sketch datasets are based on vector images of sketches not their underlying parametric relational geometry. 
For reasoning about parametric CAD sketches and inferring their design steps using deep learning models, a large-scale dataset of parametric CAD sketches is needed. 

\subsubsection{SketchGraphs \cite{seff2020sketchgraphs}}
\label{subsubsec:sketchgraphs}
The first dataset introduced in this regard is SketchGraphs, which is a collection of $15$ million real-world 2D CAD sketches from Onshape platform \cite{onshape}, each of which represented as a geometric constraint graph where the nodes are the geometric primitives and edges denote the designer-imposed geometric relationships between primitives. An open-source data processing pipeline is also released along with the dataset to facilitate further research on this area.\footnote{https://github.com/PrincetonLIPS/SketchGraphs} SketchGraphs does not only provide the underlying parametric geometry of the sketches, but it also provides ground truth construction operations for both geometric primitives and the constraints between them. Therefore, it can be used for training deep learning models for different generative applications facilitating the design process. One of these applications which could be used as an advanced feature in CAD software is to automatically build the parameteric CAD model given a hand-drawn sketch or a noisy scan of the object. 
The SketchGraphs processing pipeline can be used to produce noisy renderings of the sketches. In this way, a large-scale dataset or paired geometric sketches and their noisy rendered images can be created to train a deep learning model for predicting the design steps of a sketch given its hand-drawn image. The 3D design of the model can then be obtained by extruding the designed 2D sketch. 
In \cite{seff2020sketchgraphs}, an auto-regressive model is proposed which employs SketchGraphs dataset for two use cases: 1) Autoconstrain, which is conditional completion of an sketch by generating constraints between primitives given the unconstrained geometry, and 2) Generative modeling, which is auto-completing a partially designed sketch by generating construction operations for adding the next primitives and constraints between them. Although most of the CAD software have a built-in constraint solver to be used in design process, these generative methods are useful when an unconstrained sketch is uploaded to the software as a drawing scan designer needs to find the constraints between the sketch primitives and/or complete the sketch. 
This problem is quite similar to program synthesis or induction in constraint programming. The SketchGraphs dataset and its generative methods can be a good baseline for the future works in this direction. 
In the following, the two generative use-cases are described in more detail. 

Let us assume a sketch represented by a multi-hypergraph $\mathcal{G=(V,E)}$ where nodes $\mathcal{V}$ denote primitives and edges $\mathcal{E}$ denote constraints between them. In this graph, each edge might connect one or more nodes, and multiple edges might share the same set of connected nodes. An edge with a single node is indicated as a self-loop showing a single constraint (e.g. length) for a single primitive (e.g. line), and a hyper-edge applies on three or more nodes (e.g. mirror constraint applied on two primitives while assigning one more primitive as an axis of symmetry). An example of a sketch graph is illustrated in Figure \ref{fig:sketchgraph}.
Each constraint is identified by its type, and each primitive is identified by its type and its parameters (different primitives might have different number of parameters). 
%
For Autoconstrain task, all the graph nodes (primitives) are given, and the model is trained in a supervised manner to predict the sequence of graph edges considering the ground truth ordering of constraints in the dataset. This problem can be seen as an example of graph link prediction \cite{zhang2018link} which predicts the induced relationships between the graph nodes. 
Starting from the first construction step, the model first predicts which node should be connected to the current node and then creates a link (edge) between these two neighboring nodes. Then it predicts the type of this edge (constraint). 
For generative modeling task which auto-completes a partially completed sketch (graph) by generating new primitives (nodes) and constraints between them (edges), the primitives are only represented by their type (the primitive parameters are ignored) and the constraints are represented by both their type and their numerical or categorical parameters. As an example, if the constraint between two primitives is \textit{distance}, its parameter could be a scalar value denoting the Euclidean distance between the two primitives. 
However, while this model predicts both type and parameters for constraints, it only predicts the type of the primitives, not their parameters, and the initial coordinates of the primitives might not fit into the sketch and constraints. Therefore, the final configuration of the primitive coordinates in the sketch needs to be found by the CAD software's built-in geometric constraint solver. This issue is addressed in the next proposed work for 2D sketch generation, CurveGen-TurtleGen \cite{willis2021engineering}, which is introduced in \ref{subsubsec:curvegen_turtlegen}.

\subsubsection{CurveGen-TurtleGen \cite{willis2021engineering}}
\label{subsubsec:curvegen_turtlegen}
\begin{figure}
    \centering
    \includegraphics[width=0.5\textwidth]{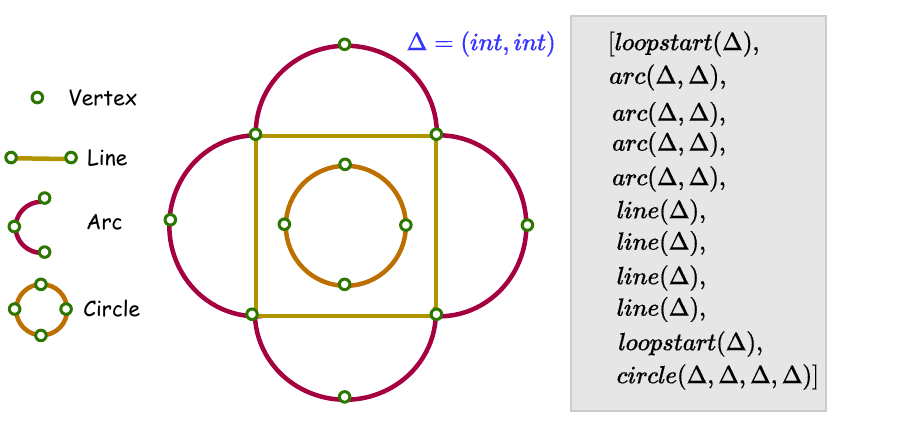}
    \caption{The illustration showcases a basic sketch composed of $12$ vertices and $9$ hyperedges of types \textit{line}, \textit{circle} and \textit{arc} entities. On the right side, a sequence of commands for rendering this sketch, following the grammar proposed by \cite{willis2021engineering}, is presented. Each \textit{line} connects $2$ vertices denoted by $\Delta$, every \textit{arc} is defined by traversing $3$ vertices, and \textit{circles} pass through $4$ vertices. Notably, in each loop, the initiation point of each entity coincides with the termination point of the preceding entity in the sequence.
    }
    \label{fig:hypersketch}
\end{figure}

As discussed previously in \ref{subsubsec:sketchgraphs}, the generative models in SketchGraphs rely on the built-in sketch constraint solver in CAD software to set the final configuration of the produced sketch. This issue is addressed by two generative models, CurveGen and TurtleGen, proposed in \cite{willis2021engineering} which encode constraint information implicitly in the geometric coordinates to be independent from the sketch constraint solver. 
In this work, the primitive types are simply limited to lines, curves, and circles, as they are the most common primitives used in 2D sketches, and it is considered that the constraints between the primitives in sketches should be defined in a way that the geometric primitives can form closed profile loops. In this regard, this method proposed two different representations for engineering sketches: Sketch Hypergraph representation which is used by CurveGen, and Turtle Graphics representation which is used by TurtleGen. 
In the Sketch Hypergraph representation, the sketch is represented as a hypergraph $\mathcal{G}=(\mathcal{V}, \mathcal{E})$, where a set of vertices $\mathcal{V} =\left \{ \nu_1, \nu_2, ..., \nu_n \right \}$ with their corresponding 2D coordinates $\nu_i = (x_i, y_i)$ are encoded as graph nodes, while $\mathcal{E}$ denotes a set of hyperedges connecting two or more vertices to make different primitives. The primitive type is defined by the cardinality of the hyperedge. For example, the primitive \textit{line} is made by two connected vertices, \textit{arc} is made by three connected vertices and \textit{circle} can be seen as a set of four connected vertices. 
Figure \ref{fig:hypersketch} shows an example of a simple sketch consisting of $12$ vertices and $9$ hyperedges. 
This representation is used by CurveGen, which is an autoregressive Transformer based on PolyGen \cite{nash2020polygen}. PolyGen is a an autoregressive generative method for generating 3D meshes using Transformer architectures which are able to capture long-range dependencies. The mesh vertices are modeled unconditionally by a Transformer, and the mesh faces are modeled conditioned on the mesh vertices by a combination of Transformers and pointer networks \cite{vinyals2015pointer}. 
Similar to PolyGen, CurveGen also generates directly the sketch hypergraph representation by first generating the graph vertices $\mathcal{V}$ which are used to make curves, and then generating the graph hyperedges $\mathcal{E}$ conditioned on generated vertices as follows: 
\begin{equation}
    p(\mathcal{G}) = p(\mathcal{E}|\mathcal{V})p(\mathcal{V}). 
    \label{eq:chainrule}
\end{equation}
In this way, the network predicts the precise coordinates of each primitive while the type of primitive is implicitly encoded in the hyperedges which group vertices together to make different types of primitives. Therefore, this method is independent of any constraint solver for finding the final primitives configuration. By having the precise coordinates of the sketch curves, the constraints between them can be automatically obtained as a post-processing step. However, implicit inference of constraints makes editing the sketch in the software more difficult. If the designer wants to change one of the constraints between the primitives, for example scale of a distance, this change will not propagate through the whole sketch and primitives, because the exact positioning of the primitives are somehow fixed. 
In the Turtle Graphics representation, the sketch is represented by a sequence of drawing commands, pen-down, pen-draw, pen-up, which can be executed to form an engineering sketch in the hypergraph representation. The TurtleGen network is an autoregressive Transfromer model generating a sequence of drawing commands to iteratively draw a series of closed loops forming the engineering sketch. 
In this way, each sketch is represented as a sequence of loops, and each loop is made by a \textit{LoopStart} command which lifts the pen, displace it to the specified position, put it down, and starts drawing parametric curves specified by the \textit{Draw} command. In Figure \ref{fig:hypersketch}, a simple sketch with its corresponding command sequence is illustrated. 
The sketch encompasses $2$ loops and $9$ \textit{Draw} commands, including \textit{arc}, \textit{line}, and \textit{circle} types. The initial loop starts at a position $\Delta = (int, int)$ and comprises $4$ arcs. Each arc originates from the endpoint of the preceding arc, passing through $2$ additional vertices. This loop transitions into a sequence of $4$ lines. The first line begins where the last arc ended, traversing one more vertex, and this pattern repeats for the subsequent lines. The second loop is a circle that begins at another position $\Delta$ and passes through $3$ additional vertices.
CurveGen and TurtleGen have been evaluated on the SketchGraphs dataset, demonstrating superior performance compared to the generative models proposed in SketchGraphs.
We refer the reader to \cite{willis2021engineering} for further details on models' architecture, training and evaluation settings.

\subsubsection{SketchGen \cite{para2021sketchgen}}
\begin{figure*}
    \centering
    \includegraphics[width=\textwidth]{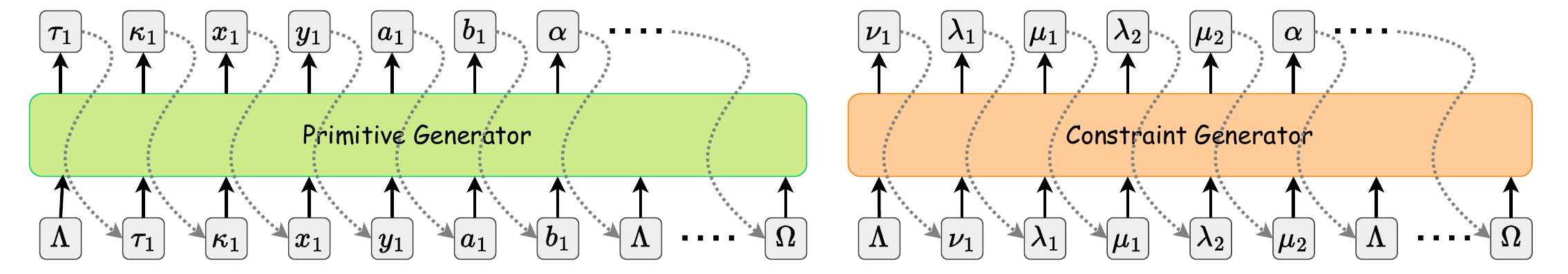}
    \caption{A simple illustration of the SketchGen generative approach, comprising two generative networks for primitive and constraint generation. At each generation step, the network produces the next token based on both the input and the previously generated token. The start and end of a new primitive or constraint sequence are denoted by $\Lambda$ and $\Omega$, respectively. The figure showcases the parameters of a line primitive and a parallelism constraint as an example. More details about the model structure can be found in the original paper \cite{para2021sketchgen}.}
    \label{fig:sketchgen}
\end{figure*}

Concurrent to CurveGen-TurtleGen, SketchGen, which is an autoregressive generative method, proposed based on PolyGen \cite{nash2020polygen} and pointer networks \cite{vinyals2015pointer}. Unlike PolyGen, SketchGen aims to capture the heterogeneity of the primitives and constraints in sketches, where each type of primitive and constraint might have a different number of parameters, each of different type, and therefore have a representation of a different size. The selection of input sequence representation has a great impact on the performance of the Transformers, and transforming heterogeneous constraint graphs of sketches into an appropriate sequence of tokens is considerably challenging. One simple solution is padding all the primitives' and constraints' representations to make them of the same size. However, this technique is inefficient and inaccurate for complex sketches. 
SketchGen proposed a language with a simple syntax to describe heterogeneous constraint graphs effectively. 
The proposed language for CAD sketches encodes the constraints and primitives parameters using a formal grammar. The terminal symbols for encoding the type and parameters of primitives are $ \left \{ \Lambda,  \Omega, \tau, \kappa, x, y, u, v, a, b \right \}$ and for constraints are $ \left \{ \Lambda, \nu, \lambda, \mu, \Omega \right \}$. The start and end of a new primitive or constraint sequence are marked with $\Lambda$ and $\Omega$, respectively. 
The primitive type is denoted with $\tau$, and $\nu$ shows the the constraint type. $\kappa, x, y, u, v, a, b$ show the specific parameters for each primitive, such as coordinate and direction. $\lambda$, and $\mu$ are the specific constraint parameters indicating the primitive reference of the constraint and the part of the primitive it is targeting, respectively. This formal language, enables distinguishing different primitive or constraint types along with their respective parameters.
For example, the sequence for a line primitive is $\Lambda, \tau, \kappa, x, y, u, v, a, b, \Omega$, which starts with $\Lambda$, followed by the primitive type $\tau$ = \textit{line}, the construction indicator $\kappa$, the coordinates of the starting point $x$ and $y$, the line direction $u$ and $v$, the line range $a$, $b$, and ends with $\Omega$.
The sequence for a parallelism constraint is $\Lambda, \nu, \lambda_1, \mu_1, \lambda_2, \mu_2, \Omega$ which starts with $\Lambda$, followed by the constraint type $\nu$ = \textit{parallelism}, the reference to the first primitive $\lambda_1$, the part of the first primitive $\mu_1$, the reference to the second primitive $\lambda_2$, the part of the second primitive $\mu_2$, and ends with $\Omega$.
In this way, each token $q_i$ (a sequence of symbols) represents either a primitive or a constraint, and each sketch $Q$ is represented as a sequence of tokens. 
Similar to PolyGen and CurveGen-TurtleGen, as stated in Eq. (\ref{eq:chainrule}), the generative model of SketchGen is also decomposed into two parts by first generating the primitives $p(\mathcal{P})$ and then generating constraints conditioned on primitives $p(\mathcal{C}|\mathcal{P})p(\mathcal{P})$ as follows: 
\begin{equation}
    p(\mathcal{S}) = p(\mathcal{C}|\mathcal{P})p(\mathcal{P}). 
    \label{eq:sketchgen}
\end{equation}
Therefore, the generative network learns the distribution of constraint sketches via two autoregressive Transformers, one for primitive generation, and the other for conditional constraint generation. 
The sketch is parsed into a sequence of tokens, by initially iterating through all the primitives and expressing them using the language sequences explained above. Subsequently, a similar process is applied to represent all the constraints within the sketch. 
As illustrated in Figure \ref{fig:sketchgen}, the input of the primitive generator network is a sequence of concatenated primitive tokens seperated by $\Lambda$, such as $\Lambda, \tau_1, \kappa_1, x_1, y_1, u_1, v_1, a_1, b_1, \Lambda, \tau_2, \kappa_2, x_2, y_2, ..., \Omega$, and the input of the constraint generator is a sequence of concatenated constraint tokens seperated by $\Lambda$, like $\Lambda, \nu_1, \lambda_{11}, \mu_{11}, \lambda_{12}, \mu_{12}, \Lambda, \nu_2, \lambda_{21}, .... \Omega$.

All the primitive and constraint parameters are quantized first and then mapped by an embedding layer to make the input feature vectors for the network. The positional information of the tokens in the sequence is also captured by a positional encoding added to each embedding vector. The resulting sequences of tokens are then fed into the Transformer architectures to generate primitives and constraints. The constraint generator network not only receives the embedded and positionally encoded tokens for constraints as input, but also it receives the embedded and positionally encoded sequence of tokens that represents the primitives generated in the previous step, to generate constraints conditioned on primitives. 
Similar to \cite{seff2020sketchgraphs}, this model is evaluated on SketchGraphs dataset for the two tasks of constraint prediction given sketch primitives, and full sketch generation from scratch by generating both primitives and constraints sequentially. The final generated sketch needs to be regularized by a constraint solver to remove the potential errors caused by quantizing the sketch parameters. 

\subsubsection{CAD as Language \cite{ganin2021computer}}
Concurrent to CurveGen-TurtleGen and SketchGen, CAD as Language is another autoregressive Transformer method proposed for 2D sketch generation based on PolyGen \cite{nash2020polygen}. Unlike CurveGen-TurtleGen that predicts only the primitive parameters with implicit constraints and independent of constraint solver, CAD as Language and SketchGen methods generate both primitives and constraints but are dependent to a built-in constraint solver to obtain the final configuration of the sketch. However, SketchGen produces primivies and constrains via two separate Transformer networks while do not support arbitrary orderings of primitive and constraint tokens. In SketchGen, the sketch is parsed into a sequence of tokens, by initially iterating through all the primitives and then iterating through all the constraints.
CAD as Language method, not only handles the arbitrary orderings of primitives and constraints, but also generates both primitives an constraint via one Transformer network. 
Unlike all the previous sketch generation methods which are evaluated on SketchGraphs dataset, CAD as Language evaluated its generative method on a new collection of over $4.7$ million sketches from the Onshape platform which avoids the problem of data redundancy in SketchGraphs. This collected dataset and the corresponding processing pipeline are publicly available on Github.\footnote{https://github.com/google-deepmind/deepmind-research/tree/master/cadl} 

\begin{figure}
    \centering
    \includegraphics[width=\linewidth]{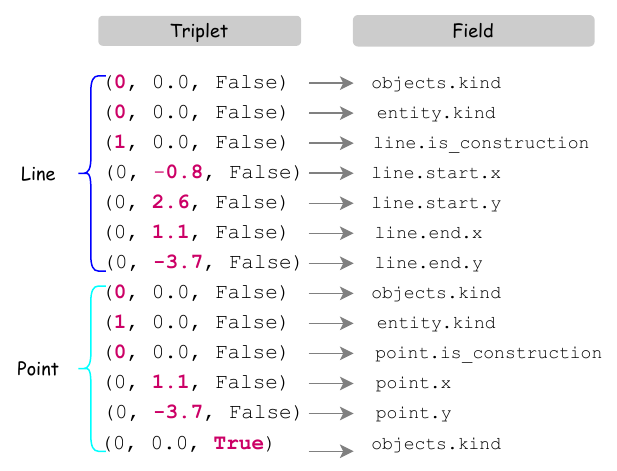}
    \caption{The description of a simple sketch, consisting of a line and one point on one of its ends, using the language structure proposed by \cite{ganin2021computer}. The active element of each triplet (on left side) is specified in bold red color, and the corresponding field of the object for each triplet is shown on the right side.}
    \label{fig:cadaslang}
\end{figure}

CAD as Language uses a method for describing structured objects using Protocol Buffers (PB) \cite{varda2008protocol}, which demonstrates more efficiency and flexibility for representing the precise structure of complex objects than JSON format. In this format, each sketch is described as a PB message. 
Similar to other Transformer-based methods, the first and most important step in the processing pipeline is to parse sketches to a sequence of tokens. In this method, each sketch (or PB message) is represented as a sequence of triplets $(d_i, c_i, f_i)$, where each triplet with index $i$ denotes a token. Each token (triplet) represents only one component (type or parameter) of a primitive or constraint in a sketch, where $d_i$ is a discrete value denoting the type of object it is referring to, the type of entity (primitive, constraint, etc), $c_i$ is a continuous value denoting the parameter value for the corresponding entity. At each time, either $d_i$ or $c_i$ is active and gets a value, and the other one is set to zero. $f_i$ is a boolean flag specifying the end of a repeated token (for example the end of an object containing an entity). 
\ref{fig:cadaslang} shows an example of tokens specifying a line and a point primitive on one of its ends, respectively. 

As it is shown in this example, the first triplet (\textit{objects.kind}) is always associated with the type of the object the token referring to. The values in the second triplet depends on the type of object specified in the first triplet. As in this example, $d_1 = 0$ shows that this sequence of tokens is about creating a primitive (like a line), therefore the second triplet specifies the type of the primitive (\textit{entity.kind)} which is $0$ for \textit{line}, and $1$ for \textit{point}. The next triplets in the sequence specify the specific parameters associated with the corresponding primitive identified in the second triplet. For example, \textit{line} primitive is defined with a start and ending point, while the \textit{point} primitive, which is also repeated in the \textit{line} primitive, is defined with $x$, $y$ coordinates. 

For interpreting these triplets (tokens), CAD as Language method also proposed a custom interpreter which receives as input a sequence of tokens, each representing a sketch component (which can be an entity type, parameter or any other design step), and converts it into a valid PB message. This interpreter is designed in a way to handle arbitrary orderings of tokens and make sure that all the token sequences can be converted into a valid PB message (sketch). 
This interpreter guides the Transformer network through the sketch generation process. The Transformer network receives as input a sequence of tokens, and at each time step it outputs a raw value which is passed to the interpreter to infer the corresponding triplet for that value. This triplet is a part of a PB message which makes the final sketch. When the interpreter inferred the output value, it propagates its interpretation back to the Transformer to guide it through generating the next value. 
Therefore, the structure this method proposed for parsing a sketch into tokens and interpreting them, enables it to generate every sketch component (primitives and constraints) via only one Transformer network, while handling different orderings of input tokens. 
A conditional variant of the proposed Transformer model is also explored and evaluated in this method, where it is conditioned on an input image of the sketch.

\subsubsection{VITRUVION \cite{SeffZRA22}}

This method is the latest autoregressive generative model proposed for sketch generation. Similar to SketchGen and CAD as Language, VITROVION also generates both primitives and constraints autoregressively. 
However, this method is more similar to SketchGen in a sense that primitives and constraints are generated independently through training two distinct Transformer networks. 
The major contribution of this method compared to the previous ones, is conditioning the model on various contexts such as hand-drawn sketches. This contribution is one step forward towards the highly-sought feature in CAD softwares to reverse engineer a mechanical part given a hand-drawing or a noisy scan of it.  
Generating parametric primitives and constraints of a sketch given a hand-drawing or noisy image saves a lot of time and effort in designing process. 

Similar to the previous methods, VITROVION is a generalization of PolyGen \cite{nash2020polygen}. It generates propability distribution of sketches by first generating primitives $\mathcal{P}$, and then generating constraints $\mathcal{C}$ conditioned on primitives. However, in this method, the primitive generation is optionally conditioned on an image follows: 
\begin{equation}
    p(\mathcal{P}, \mathcal{C} | I) = p(\mathcal{C}|\mathcal{P})p(\mathcal{P}|\mathcal{I}), 
    \label{eq:vitrovion}
\end{equation}
where $\mathcal{I}$ is a context such as hand-drawn image of a sketch. However, constraint modeling in this method only supports constraints with one or two reference primitives, and not hyperedges connecting more than two primitives as in CurveGen-TurtleGen method. 
The generative models are trained and evaluated on SketchGraphs dataset for autoconstraint, autocomplete, and conditional sketch synthesis tasks. 
As explained in \ref{subsubsec:sketchgraphs}, in autoconstraint application the network generates constraints conditioning on a set of available primitives, but in autocomplete task, an incomplete sketch is completed by generating both primitives and constraints. In both cases, the constraint generator network is conditioned on generated primitives which can be imperfect. VITROVION increases the robustness of the constraint generator network by conditioning it on noise-injected primitives. 
The final primitive and constraint parameters are finally adjusted via a standard constraint solver.
For image-conditional sketch synthesis, the model infers  primitives conditioning on a raster-image of a hand-drawn sketch. In this regard, an encoder network based on Vision Transformer \cite{dosovitskiy2020image} architecture is used to obtain the embeddings of the image patches, and the primitive generator network then cross-attends to these patch embeddings for predicting sketch primitives. 
This idea is based on a similar idea in PolyGen for image-conditioned mesh generation. 

This method tokenizes a sketch by representing each primitive and constraint as a tuple of three tokens: value, ID, position. 
The value token has two-folds, one indicating the type of primitive or constraint and the other indicating the numerical value of the associated parameter of that primitive or constraint. ID token indicates the type of parameter specified by the value token, and position token indicates the ordered index of the primitive or constraint to which this ID and value tokens belong to. 
The ordering of primitives are according to design steps indicated in SketchGraphs dataset, and the ordering of constraints are according to the ordering of their corresponding reference primitives.

\subsection{3D CAD Generation from Sketch}


\subsubsection{Sketch2CAD \cite{li2020sketch2cad}} The first work proposed for sequential 3D CAD modeling by interactive sketching in context is Sketch2CAD. This work is a learning-based interactive modeling system that unifies CAD modeling and sketching by interpreting the user's input sketches as a sequence of parametric 3D CAD operations. 
Given an existing incomplete 3D shape and input sketch strokes added on it by the user, Sketch2CAD first obtains the normal and depth maps of sketching local context which are introduced to the CAD operator classification and segmentation networks. 
Classification is done by a CNN network predicting the type of CAD operation needed for creating the corresponding input sketch on the 3D shape. It receives as input the concatenation of three maps, each of size $256 \times 256$, i.e., the sketching map representing stroke pixels with binary values, and normal and depth maps representing the local context via rendering the shape through a specific viewpoint. 
It should be noted that only four types of CAD operations, which are the most widely used ones, are supported in this method, i.e., \textit{Extrude, Add/Subtract, Bevel, Sweep}. According to the predicted operation type, the parameters of the operator are regressed via specific segmentation networks. For example, for the predicted \textit{Sweep} operator, a SweepNet is trained to infer the corresponding parameters. Each of the four segmentation networks have U-Net structure with one encoder and two decoders, one producing the probability map of base face for the sketch and the other one producing the corresponding curve segmentation map. The segmentation is followed by an optimization process for fitting the operation parameters. 

The parametric nature of the recognized operations provides strong regularization to approximate input strokes and allows users to refine the result by adjusting and regressing parameters. The modeling system also incorporates standard CAD modeling features such as automatic snapping and autocompletion. The output of the system is a series of CAD instructions ready to be processed by downstream CAD tools. 
Since no dataset of paired sketches and 3D CAD modeling operation sequences exists, Sketch2CAD introduced a synthetic dataset of $40,000$ shapes for training and $10,000$ shapes for testing, containing step-by-step CAD operation sequences and their corresponding sketch image renderings. The dataset, code, and visual examples are publicly available.\footnote{https://geometry.cs.ucl.ac.uk/projects/2020/sketch2cad/}



\subsubsection{Free2CAD \cite{li2022free2cad}} 
One of the main challenges of Sketch2CAD is that it can only handle one CAD operation at a time, which means that it assumes that at each step, the user drawing added to the shape corresponds to only one CAD operation. 
It requires the user to draw the sketch sequentially and part-by-part so that it can be decomposed into the meaningful CAD operations. 
Free2CAD is proposed to address this restriction. This is the first sketch-based modeling method in which the user can input the complete drawing of a complex 3D shape without needing to have expert knowledge of how this sketch should be decomposed into the CAD operations or following a specific strategy for drawing the sketch and working with the system. 

Free2CAD is a sequence-to-sequence Transformer network which receives as input a sequence of user drawn strokes depicting a 3D shape, processes and analyzes them to produce a valid sequence of CAD operation commands, which may be executed to create the CAD model. The main contribution of this method is the automatic grouping of the sketch strokes and the production of the parametric CAD operations for each group of strokes sequentially, conditioned on the groups that have been reconstructed in the previous iterations. 
The method is comprised of two phases, namely stroke grouping phase and operation reconstruction phase. 
In the stroke grouping phase, first each sketch stroke is embedded as a token via a specially designed Transformer encoder network, then it is processed by the Transformer decoder network which produces group probabilities for the input tokens. In this way, the sketch strokes which might make a specific part of the shape are grouped together. 
The most closely related work to the stroke grouping phase is the SketchGNN \cite{yang2021sketchgnn} method which proposes a Graph Neural Network approach for freehand sketch semantic segmentation. 
Next, in the operation reconstruction phase, the candidate groups are converted into geometric primitives with their corresponding parameters, conditioned on the existing geometric context. This step is followed by geometric fitting and grouping correction before passing back the updated groups as geometric context for the next iteration. 
At the end of this process, the desired CAD shape and the sequence of CAD commands are obtained.
The method is also extended to handle long stroke sequences making complex shapes using a sliding window scheme progressively outputting CAD models. 

Similar to the Sketch2CAD \cite{li2020sketch2cad}, Free2CAD also provides a large-scale synthetic dataset of $82,000$ paired CAD modeling operation sequences and their corresponding rendered sketches which are segmented based their corresponding CAD commands. The code and dataset of this method are publicly available \footnote{https://geometry.cs.ucl.ac.uk/projects/2022/free2cad/}.
The evaluation results of Free2CAD on both their generated dataset and on Fusion 360 dataset illustrate its high performance in processing different user drawings and producing CAD commands which make desirable 3D shapes when executed by CAD tools. 


\subsubsection{CAD2Sketch \cite{hahnlein2022cad2sketch}}
\begin{figure*}
    \centering
    \includegraphics[width=0.8\textwidth]{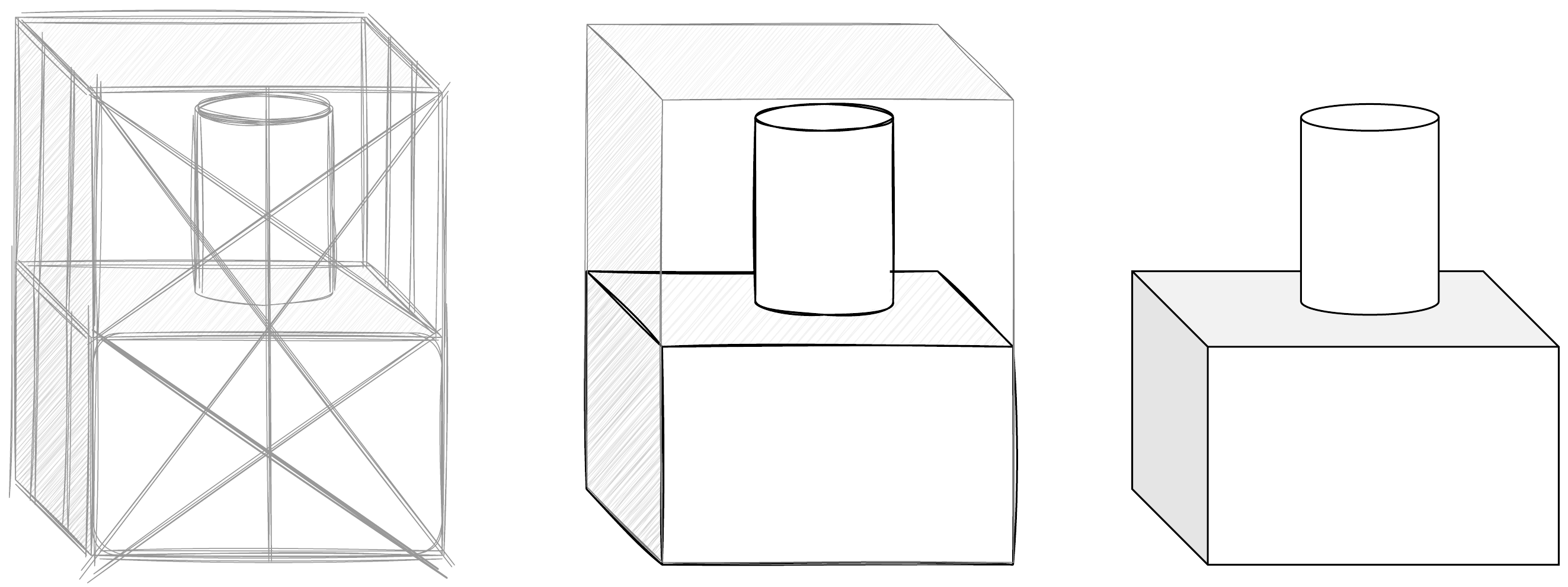}
    \caption{An illustration showcasing a conceptual sketch (left) designed for creating a 3D shape. Examples of these sketches can be found in the CAD2Sketch dataset \cite{hahnlein2022cad2sketch}. After refining the concept sketch, the final freehand sketch of the model is achieved (right). Refined sketches like these are available in the Sketch2CAD dataset \cite{li2020sketch2cad}.}
    \label{fig:conceptsketch}
\end{figure*}

Unlike Sketch2CAD which tries to facilitate the design process for non-expert users in an interactive modeling system, CAD2Sketch is designed to assist expert industrial designers. As the name suggests, CAD2Sketch is a method dedicated to synthesizing concept sketches from CAD models. Concept sketching is a preliminary stage in CAD modeling, wherein designers refine their mental conception of a 3D object from rough outlines to intricate details, often using numerous construction lines. An example of a simple concept sketch is depicted in \ref{fig:conceptsketch}, shown on the left side, alongside its refined version displayed on the right side.
Notably, concept sketching mirrors the detailed steps in a designer's mind, akin to the stages of CAD modeling, while sketches in the Sketch2CAD dataset typically present the final free-hand sketch without these detailed auxiliary construction lines. 
CAD2Sketch introduced a large-scale synthetic dataset of concept sketches by proposing a method for converting CAD B-Rep data into concept sketches. CAD2Sketch establishes a large-scale synthetic dataset of concept sketches by introducing an method to convert CAD B-Rep data into concept sketches. The dataset is comparable to the OpenSketch \cite{gryaditskaya2019opensketch} dataset which has $400$ real concept sketches crafted by various expert designers. However, CAD2Sketch targets to bridge the gap between synthetic and real data, enabling the training of neural networks on concept sketches.
The CAD2Sketch method initially generates construction lines for each operation in a CAD sequence. To avoid overwhelming the sketch with too many lines, a subset of these lines is chosen by solving a binary optimization problem. Subsequently, the opacity and shape of each line are adjusted to achieve a visual resemblance to real concept sketches.
The synthetic concept sketches produced by CAD2Sketch closely resemble their corresponding real pairs in OpenSketch, so that even designers can hardly distinguish between them \cite{hahnlein2022cad2sketch}. CAD2Sketch also generates a substantial number of paired sketches and normal maps, utilized for training a neural network to infer normal maps from concept sketches.
The dataset contains approximately $6,000$ paired concept sketches and normal maps. The trained neural network's generalization to real shapes is evaluated on a test set of $108$ CAD sequences from the ABC dataset, yielding promising results.
However, it is worth noting that this method is evaluated on a limited number of real sketches, given the relatively small size of the OpenSketch dataset. Additionally, the CAD2Sketch method, built on CAD sequences from existing large-scale CAD datasets, is limited to sequences composed of sketch and extrusion operations.

\subsection{3D CAD Command Generation}

While previous approaches have introduced synthetic datasets for 3D CAD reconstruction, the absence of a standardized collection of human-designed 3D CAD models with retained CAD command sequences poses a limitation. Similar to the valuable contribution of the SketchGraphs dataset in the area of 2D sketch synthesis, a curated dataset of human-designed 3D CAD models with preserved CAD command sequences would greatly benefit research and the development of practical methods for real-world applications. 
Fusion 360 Reconstruction dataset \cite{willis2021fusion} fills this gap as the first human-designed dataset, featuring $8,625$ CAD models constructed using a sequence of \textit{Sketch} and \textit{Extrude} CAD operations. Accompanying this dataset is an environment known as the Fusion 360 Gym, capable of executing these CAD operations. Each CAD model is represented as a Domain-Specific Language (DSL) program, a stateful language serving as a simplified wrapper for the underlying Fusion 360 Python API. This language keeps track of the current geometry under construction, updated iteratively through a sequence of \textit{sketch} and \textit{extrude} commands. 
The data and correponding codes are publicly available \footnote{https://github.com/AutodeskAILab/Fusion360GalleryDataset/blob/master/docs/reconstruction.md}.
This dataset is benchmarked through the training and evaluation of a machine learning-based approach featuring neurally guided search for programmatic CAD reconstruction from a specified geometry. The approach begins by training a policy, which is instantiated as a Message Passing Network (MPN) \cite{duvenaud2015convolutional, gilmer2017neural} with an original encoding of state and action. This training is conducted through imitation learning, drawing insights from ground truth construction sequences. In the subsequent inference stage, the method incorporates a search mechanism, utilizing the learned neural policy to iteratively engage with the Fusion 360 Gym environment until a precise CAD program is discerned. However, a limitation of this method is its assumption that the 2D geometry is given, with the method solely predicting the sketches to be extruded and their extent at each iteration.
Subsequently, this dataset has been utilized by generative methods that tackle scenarios where the geometry is not provided, and the entire 3D model needs to be synthesized from scratch using \textit{Sketch} and \textit{Extrude} operations. 

As another effort to infer CAD modeling construction sequences through neural-guided search, a method in which the B-Rep data of each CAD model is expressed in the form of a zone graph, where solid regions constitute the zones and surface patches (curves) form the edges is proposed in \cite{xu2021inferring}. This representation is then introduced to a GCN to facilitate feature learning and, subsequently, a search method is employed to deduce a sequence of CAD operations that can faithfully recreate this zone graph. 
However, none of these two aforementioned methods leverage generative methods for CAD operation generation.
In this subsection, the most recent generative methods in this domain are introduced.

\subsubsection{DeepCAD \cite{wu2021deepcad}}
The first generative deep learning model proposed for creating 3D CAD commands is DeepCAD. 
Given the sequential and irregular nature of CAD operations and B-Rep data, it is essential to choose a fixed set of the most commonly employed CAD operations and organize them into a unified structure for utilization by a generative neural network. Drawing inspiration from earlier generative techniques designed for 2D CAD analysis and sketch generation such as CurveGen-TurtleGen, SketchGen, CAD as Language, and VITROVION outlined in Section \ref{subsec:2dsketchgen}, DeepCAD follows a similar approach by likening CAD operations to natural language. It introduces a generative Transformer network for autoencoding CAD operations. It is worth noting that DeepCAD diverges from previous generative methods by adopting a feed-forward Transformer structure instead of the autoregressive Transformer commonly used in such contexts.
Additionally, DeepCAD has constructed and released a large-scale dataset featuring $178,238$ CAD models from Onshape repository generated through \textit{Sketch} and \textit{Extrude} operations, accompanied by their respective CAD construction sequences. This dataset surpasses the size of the Fusion 360 Gallery dataset, which contained approximately $8,000$ designs. The substantial increase in the number of designs within the DeepCAD dataset enhances its suitability for training generative networks.

\begin{figure}
    \centering
    \includegraphics[width=\linewidth]{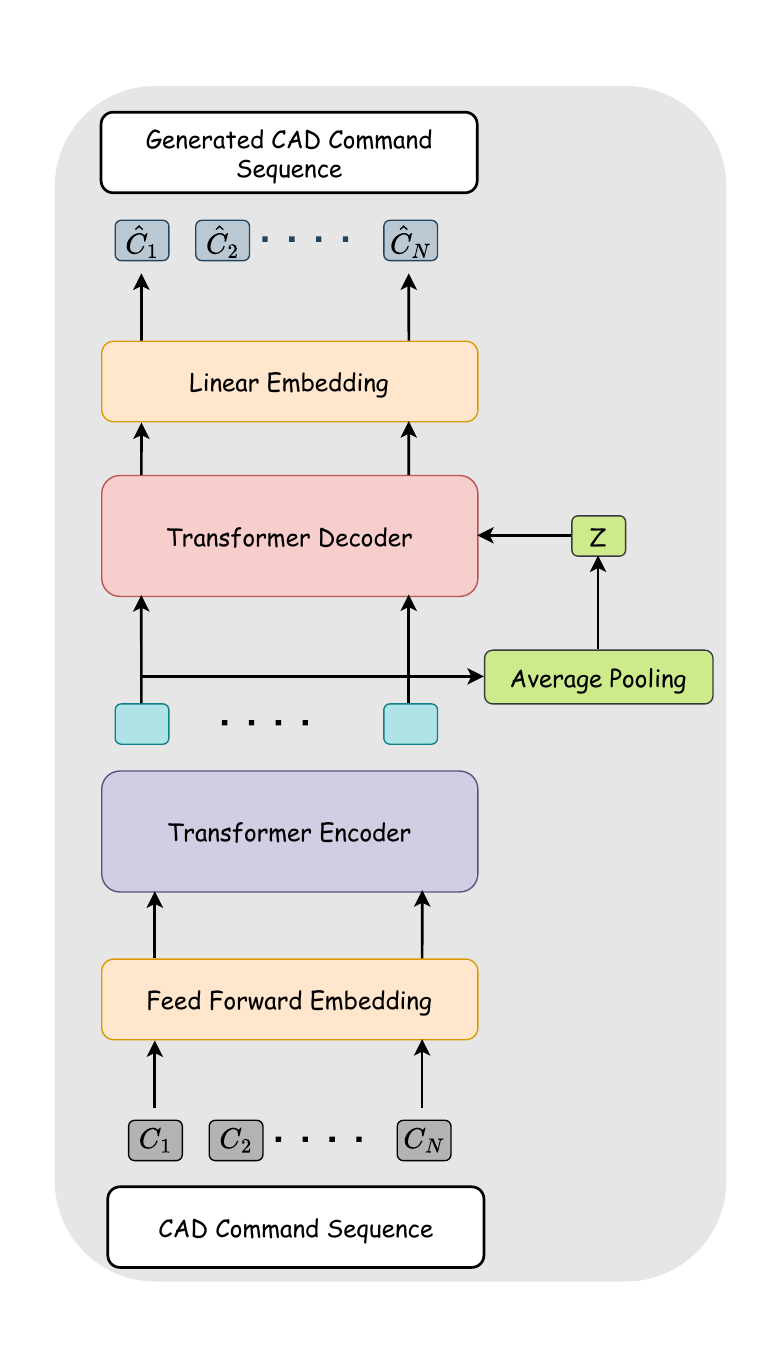}
    \caption{Schematic representation of the DeepCAD model architecture \cite{wu2021deepcad}. The model, is a transformer autoencoder network trained in an unsupervised manner. It takes a sequence of CAD commands $\left \{ C_1, C_2, ..., C_N \right \}$ as input and generates the reconstructed commands $\left \{\hat{C}_1, \hat{C}_2, ..., \hat{C}_N \right \}$ as output.}
    \label{fig:deepcad}
\end{figure}

In the standardized structure suggested for CAD operations in DeepCAD, the CAD commands are explicitly detailed, providing information on their type, parameters, and sequential index. The \textit{Sketch} commands encompass curves of types \textit{line}, \textit{arc} and \textit{circle} along with their respective parameters. Meanwhile, \textit{Extrude} commands signify extrusion operations of types \textit{one-sided, symmetric, two-sided} and boolean operations of types \textit{new body, join, cut, or intersect}. These operations are employed to integrate the modified shape with the previously constructed form.
A CAD model $M$ is represented as a sequence of curve commands which build the sketch, intertwined with extrude commands. The total number of CAD commands to represent each CAD model is fixed to $60$ and a padding approach with empty commands is used to fit the CAD models with shorter command sequence in this fixed-length structure. 
Each command undergoes encoding into a $16$-dimensional vector representing the whole set of parameters of all commands. In instances where specific parameters for a given command are not applicable, they are uniformly set to $1$.
As illustrated in Figure \ref{fig:deepcad}, the autoencoder network takes a sequence of CAD commands as input, transforms them into a latent space through Transformer encoder, and subsequently decodes a latent vector to reconstruct a sequence of CAD commands. The generated CAD commands can be imported into CAD software for final user editing. 

The performance of DeepCAD is evaluated on two tasks: CAD model autoencoding and random CAD model generation. 
Once the autoencoder network is trained for reconstructing CAD commands, the latent-GAN technique \cite{achlioptas2018learning} is employed to train a generator and a discriminator on the learned latent space. The generator produces a latent vector $z$ by receiving as input a random vector sampled from a multivariate Gaussian distribution. This latent vector can then be introduced to the trained Transformer decoder to produce the CAD model commands.  
Their experiments also demonstrate that the pretrained model on the DeepCAD dataset exhibits good generalization capabilities when applied to the Fusion 360 dataset. Notably, these two datasets originate from distinct sources, namely Onshape and Autodesk Fusion 360.

The application of DeepCAD generative model is further demonstrated in the conversion of 3D point cloud data into a CAD model. In this context, both the generative autoencoder and the PointNet++ encoder \cite{qi2017pointnet} are trained concurrently to encode the CAD model into the same latent vector $z$ from the CAD commands sequence and its corresponding point cloud data, respectively. During inference, the pretrained PointNet++ \cite{qi2017pointnet} encoder embeds the point cloud data into the latent vector $z$ which is subsequently input to the generative decoder of the pretrained DeepCAD autoencoder to produce the CAD commands sequence.

\subsubsection{SkexGen \cite{skexgen22}} 
Despite advancements achieved by DeepCAD which excels in generating a diverse range of shapes, a persistent challenge remains in the limited user control over the generated designs. The ability for users to exert influence over the output and tailor designs to meet specific requirements would be a great advantage for the real world applications. 
In response to this challenge, SkexGen proposed a novel autoregressive Transformer network with three separate encoders capturing the topological, geometric and extrusion variations in the CAD command sequences separately. This approach enables more effective and distinct user control over the topology and geometry of the model, facilitating exploration within a broader search space of related designs which in turn, results in the generation of more realistic and diverse CAD models.

Inspired by CurveGen-TurtleGen \cite{willis2021engineering} and DeepCAD \cite{wu2021deepcad} methods, a CAD model in SkexGen is represented by a hierarchy of primitives with a sketch-and-extrude construction sequence. Within this hierarchy, a 3D model is composed of 3D solids, where each \textit{solid} is defined as an extruded sketch. A \textit{sketch} constitutes a collection of faces, and each \textit{face} represents a 2D surface enclosed by a \textit{loop}. A \textit{loop} is formed by one or multiple curves, including \textit{lines}, \textit{arcs} or \textit{circles}, and \textit{curve} represents the fundamental level of the hierarchy. Consequently, CAD models are encoded using five types of tokens for input to the Transformer: 1) topology tokens, denoting the curve type, 2) geometry tokens, specifying 2D coordinates along the curves, 3) end of primitive tokens, 4) extrusion tokens, indicating parameters of extrusion and Boolean operations, and 5) end of sequence tokens. 
The autoregressive Transformer network introduced by SkexGen is composed of two independent branches, each trained separately: 1) the \textit{Sketch} branch is composed of two distinct encoders dedicated to learning topological and geometrical variations in sketches. Additionally, a single decoder is employed, receiving concatenated topology and geometry encoded codebooks as input and predicting the sketch subsequence autoregressively.  
2) The “Extrude” branch comprises an encoder and decoder specifically designed to learn variations in extrusion and Boolean operations. 
Furthermore, an additional autoregressive decoder is positioned at the end, tasked with learning an effective combination of geometrical, topological, and extrusion codebooks, thereby generating CAD construction sequences. An additional autoregressive decoder on top learns an effective combination of geometrical, topological and extrusion codebooks as CAD construction sequences. 
This intricate tokenization and multifaceted network architecture allows for a nuanced and comprehensive control over both topology and geometry, enabling the generation of CAD models by effectively capturing various design aspects.

Given capacity of SkexGen to independently encode and generate \textit{Sketch} and \textit{Extrude} command sequences, this approach is versatile and can be applied to both 2D sketch generation and 3D CAD generation tasks. Evaluation results highlight its proficiency in generating more intricate designs compared to CurveGen-TurtleGen \cite{willis2021engineering} and DeepCAD \cite{wu2021deepcad} methods. Notably, SkexGen excels in supporting multi-step \textit{Sketch} and \textit{Extrude} sequences, a capability lacking in DeepCAD, which primarily produces single-step results. 



\subsubsection{Diffusion-CAD \cite{zhang2025diffusion}}
Diffusion-CAD \cite{zhang2025diffusion} introduces a diffusion-based generative model for synthesizing 3D CAD construction sequences with a particular emphasis on fine-grained user control and structural validity. In contrast to prior methods such as DeepCAD, and SkexGen, which either lack explicit controllability or rely on discrete sampling from latent codebooks, Diffusion-CAD formulates CAD command generation as a continuous denoising diffusion process over embedded command sequences. This approach enables greater control and a smoother sampling landscape for generation.

In Diffusion-CAD, each CAD model is represented as a fixed-length sequence of $N = 60$ commands, where each command consists of a type identifier (e.g., \textit{line}, \textit{arc}, \textit{circle}, \textit{extrude}) and a set of up to 17 geometric parameters. The commands are embedded into continuous vectors and distorted with Gaussian noise across time steps in the forward diffusion process. A Transformer-based denoising network is then trained to recover the original command vectors, which are finally projected back into discrete command tokens. This framework allows for generating valid and diverse CAD models from noise, while also supporting several control mechanisms during generation.
A key contribution of Diffusion-CAD is its support for multiple forms of user-defined control, including:
\begin{itemize}
    \item \textbf{Command-type control}: constraining the generation to specific types of primitives (e.g., only circles and extrusions),
    \item \textbf{Dimension control}: fixing specific geometric parameters such as radii or angles,
    \item \textbf{Partial sketch completion}: completing full CAD sequences given a user-provided prefix,
    \item \textbf{Structural control}: enforcing relationships like symmetry, perpendicularity, or collinearity via constraint-satisfaction modules,
    \item \textbf{Class-conditional generation}: generating shapes from specific semantic classes using classifier-guided diffusion.
\end{itemize}
Compared to DeepCAD \cite{wu2021deepcad}, which struggles with structural consistency, and SkexGen \cite{skexgen22}, which lacks fine control over geometry, Diffusion-CAD demonstrates more robust and flexible generation. It marks a significant step forward in CAD sequence generation, combining the strengths of deep generative modeling with explicit, user-guided controllability to produce valid and expressive CAD models.

\subsection{3D CAD Generation with Direct B-Rep Synthesis}

Creating 3D CAD designs through either the generation of CAD commands or direct B-Rep synthesis comes with a set of advantages and disadvantages.
The generative methods introduced in previous works, namely Fusion 360 Reconstruction, DeepCAD, and SkexGen, share a common goal of producing 3D CAD models by generating sequences of 3D CAD operations or commands. These commands are then processed by a solid modeling kernel in a CAD tool to recover the final CAD design in B-Rep format. Generating CAD commands, as opposed to directly creating the B-Rep, offers several advantages. Converting CAD commands into B-Rep format is feasible, while the reverse process is more challenging due to the potential ambiguity where different command sequences may result in the same B-Rep. Additionally, CAD commands are more human-interpretable, enabling users to edit designs for various applications by processing the commands in a CAD tool.
However, training models for such tasks necessitates large-scale CAD datasets that retain the history of CAD modeling operations. Consequently, datasets like DeepCAD (comprising around 190,000 models) and Fusion 360 Reconstruction (with approximately 8,000 models) are exclusively constructed for this purpose. In contrast, most large-scale datasets in the field, such as ABC (with over 1 million models), solely provide B-Rep data without a stored sequence of CAD modeling operations.
While generating CAD commands offers increased flexibility, interoperability, and user control, an alternative strategy is to directly synthesize the B-Rep. This approach may prove beneficial when leveraging existing large-scale datasets. Moreover, direct B-Rep synthesis allows for the support of more complex curves and surfaces in creating 3D shapes. This stands in contrast to CAD command generative methods, which are constrained to CAD models constructed using \textit{Sketch} and \textit{Extrude} commands with a limited list of supported curve types.

\subsubsection{SolidGen \cite{solidgen23}} 
SolidGen \cite{solidgen23} introduces a method for direct B-Rep synthesis, eliminating the necessity for a history of CAD command sequences. The approach leverages pointer networks \cite{vinyals2015pointer} and autoregressive Transformer networks to learn B-Rep topology and progressively predict \textit{vertices}, \textit{edges}, and \textit{faces} individually.
Running in parallel with SolidGen, the work by \cite{wang2022neural} is focused on 3D face identification within B-Rep data, given a single 2D line drawing. Drawing inspiration from pointer networks, this approach also employs an autoregressive Transformer network to identify edge loops in the 2D line drawing, and predicts one co-edge index at a time, corresponding to the actual planar and cylindrical faces in the 3D design.
However, SolidGen stands out as a more advantageous method. It goes beyond edge loop identification and synthesizes the complete B-Rep data for the 3D shape. Additionally, it supports the representation of all types of faces in the design, providing a more comprehensive and versatile solution.
SolidGen also introduces the Indexed Boundary Representation (Indexed B-Rep) to represent B-Reps as numeric arrays suitable for neural network use. This indexed B-Rep organizes B-Rep vertices, edges, and faces in a clearly defined hierarchy to capture both geometric and topological relations. Structurally, it consists of three lists, denoted as ${\mathcal{V}, \mathcal{E}, \mathcal{F}}$, representing \textit{Vertices}, \textit{Edges}, and \textit{Faces}, respectively. In such hierarchical structure, edges $\mathcal{E}$ are denoted as list of indices referring to vertices $\mathcal{V}$, and each face in $\mathcal{F}$ indicates an index list referring into edges $\mathcal{E}$. 
The proposed autoregressive network progressively predicts the B-Rep tokens through training three distinct Transformers for generating \textit{vertices}, \textit{edges}, and \textit{faces}, respectively. Formally, its goal is to learn a joint distribution over B-Rep $\mathcal{B}$: 
\begin{equation}
     p(\mathcal{B}) = p(\mathcal{V}, \mathcal{E}, \mathcal{F}),
\end{equation}
factorized as: 
\begin{equation}
    p(\mathcal{B}) = p(\mathcal{F}\mid \mathcal{E},\mathcal{V})p(\mathcal{E} \mid \mathcal{V})p(\mathcal{V}).
\end{equation}
This structure also allows for conditioning the distribution on an external context $c$, such as class labels, images, and voxels:  
\begin{equation}
    p(\mathcal{B}) = p(\mathcal{F}\mid \mathcal{E},\mathcal{V},c)p(\mathcal{E} \mid \mathcal{V}, c)p(\mathcal{V} \mid c).
\end{equation}

Following training, the generation of an Indexed B-Rep involves sampling \textit{vertices}, \textit{edges} conditioned on \textit{vertices}, and \textit{faces} conditioned on both \textit{edges} and \textit{vertices}. Subsequently, the obtained Indexed B-Rep can be transformed into the actual B-Rep through a post-processing step. For a more comprehensive understanding of these processes, interested readers are encouraged to refer to the original paper \cite{solidgen23}.
The efficacy of this method is assessed using a refined version of the DeepCAD dataset. Additionally, SolidGen introduces the Parametric Variations (PVar) dataset, purposefully designed for evaluating the model's performance in class-conditional generation tasks. This synthetic dataset comprises $120,000$ CAD models distributed across $60$ classes.

\subsubsection{CADParser \cite{zhou2023cadparser}}
In conjunction with SolidGen, CADParser has been introduced to predict the sequence of CAD commands from a given B-Rep CAD model. Unlike previous approaches that often utilized synthetic CAD datasets or relied on datasets like DeepCAD and Fusion 360 reconstruction, which were restricted to CAD models created with only two operations, namely \textit{Sketch} and \textit{Extrude}, CADParser has introduced a comprehensive dataset featuring $40,000$ CAD models. These models incorporate a broader range of CAD operations, including \textit{Sketch}, \textit{Extrusion}, \textit{Revolution}, \textit{Fillet}, and \textit{Chamfer}. This dataset provides a more diverse collection of CAD models constructed with five distinct types of CAD operations compared to previous datasets limited to only two operations. Each CAD model in this dataset is accompanied by both B-Rep data and the corresponding construction command sequence. 

CADParser also introduces a deep neural network architecture, referred to as the deep parser, designed to predict the CAD construction sequence for each B-Rep model. Drawing inspiration from UV-Net and BRepNet, both discussed in Sec. \ref{sec:representation_learning} as pioneering methods in representation learning for B-Rep data, CADParser treats each CAD model as a graph $\mathcal{G}=(\mathcal{V},\mathcal{E})$ where the nodes represent \textit{faces}, \textit{edges}, and \textit{coedges} of the model, and 
$\mathcal{E}$ signifies the connections between the graph nodes. The BRepNet architecture serves as the graph encoder backbone, taking node features and the constructed adjacency matrix as input, and extracting local and global features of the graph through graph convolutions and topological walks. Simultaneously, the sequence of CAD commands $S=(C_1, C_2, ..., C_n)$ that constructs the CAD model is encoded as feature vectors. These are combined with the graph-encoded global and local features and fed into a Transformer decoder to autoregressively predict the next command sequence. Similar to previous methods, CAD commands are tokenized by representing each command $C_i$ as a tuple of command type $t_i$ and command parameter $p_i$. 
$t_i$ is encoded as a $12$-dimensional one-hot vector, representing 12 different command types one at a time, and $p_i$ is a $257$-dimensional vector. This vector represents the quantized $256$-dimensional parameter vector of the command and a $1$-dimensional index indicating whether this command is used for the corresponding CAD model or not. The length of the CAD command sequence for all CAD models is fixed at $32$ for simplicity, with unused commands for each CAD model indicated by an index set to $-1$. The Transformer decoder has two separate output branches for predicting the CAD command type $t_i$ and parameter $p_i$ vectors, respectively.
For more details on the model architecture and training process, additional information can be found in \cite{zhou2023cadparser}. The contribution of this work represents a step forward in generating more diverse CAD models by incorporating various CAD commands.

\subsubsection{BrepGen \cite{xu2024brepgen}}
BrepGen \cite{xu2024brepgen} introduces a diffusion-based generative model that directly synthesizes complete B-Rep CAD models, addressing the limitations of prior methods such as SolidGen, which often lacked geometric expressiveness and struggled with topological validity. BrepGen departs from the autoregressive modeling used in SolidGen and instead employs a structured latent geometry representation, where the entire B-Rep is organized as a hierarchical tree: solids contain faces, faces contain edges, and edges contain vertices. Each node in this tree includes both topological and geometric features, which are encoded via latent embeddings.

One of the key innovations in BrepGen is its approach to handling topology. Rather than explicitly generating adjacency matrices, it duplicates topological elements across branches of the tree and reconstructs watertight solids by merging equivalent nodes. This implicit treatment simplifies the generative process but relies on post-hoc merging, which may occasionally lead to inconsistencies or self-intersections.
On the geometric side, BrepGen supports curved and freeform surfaces by encoding geometry through UV-sampled surface and curve point clouds. These are compressed via variational autoencoders (VAEs), allowing for a more compact representation of complex geometry than SolidGen, which is limited to simpler surfaces like planes and cylinders. Each part of the geometry (vertices, edges, and faces) is generated through separate denoising diffusion models, enabling expressive and controllable shape synthesis.

BrepGen demonstrates high performance on DeepCAD and ABC datasets, outperforming both DeepCAD’s sketch-based approach and SolidGen in terms of geometric integrity (e.g., Chamfer distance), uniqueness, novelty, and topological validity. It also supports conditional generation tasks such as shape interpolation and class-guided synthesis.
However, the reliance on latent geometry codes and implicit topology inference introduces certain limitations, such as reduced interpretability and dependency on successful merging of duplicated elements. These limitations are addressed by subsequent method, DTGBrepGen \cite{li2025dtgbrepgen}, which explicitly decouple topology and geometry generation to ensure structural consistency and improve control over the synthesis process.

\subsubsection{DTGBrepGen \cite{li2025dtgbrepgen}}
DTGBrepGen \cite{li2025dtgbrepgen} builds upon the strengths of BrepGen \cite{xu2024brepgen} by addressing its key limitations, specifically, the implicit and post-hoc treatment of B-Rep topology. Instead of jointly modeling geometry and topology within a latent tree structure, DTGBrepGen introduces a principled decoupling of topology and geometry into two separate stages, resulting in higher structural validity and more accurate geometric outputs.

The first stage of DTGBrepGen focuses entirely on topology generation, constructing valid edge-face and edge-vertex adjacency matrices using two Transformer-based encoder-decoder networks. This step ensures compliance with essential topological constraints, such as each edge connecting two distinct vertices and forming closed loops within each face. Unlike BrepGen, which infers these relationships indirectly through merging duplicated nodes, DTGBrepGen generates them explicitly and in a controlled manner.
Once a valid topological structure is generated, the second stage of the pipeline uses Transformer-based diffusion models to sequentially generate geometric attributes: vertex coordinates, edge curves, and face surfaces. Geometries are represented using cubic and bicubic B-splines, enabling mathematically precise modeling of complex curves and surfaces. This choice avoids the need for surface sampling and point cloud-based encoders used in BrepGen, improving both integrity and interpretability.

DTGBrepGen achieves superior performance across multiple benchmarks, on datasets such as DeepCAD and ABC.
It also supports class-conditional and point cloud-conditioned generation, showcasing its generalizability across design tasks.
By generating valid topologies first and conditioning geometry on them, DTGBrepGen produces watertight, structurally consistent CAD models with more accurate and editable geometries. This modular design also enhances interpretability and lays the groundwork for future extensions involving hierarchical modeling or integration with semantic constraints.

\subsection{3D CAD Generation from Point Cloud}
\label{sec:cad2pc}
As highlighted in the overview of existing 3D CAD generative models, the reverse engineering of CAD shapes has primarily been explored through methods utilizing either CAD sketches, B-Rep data or sequences of CAD commands as input.
These approaches have made promising advances so far in reconstructing and generating CAD models. However, an equally crucial aspect is generating CAD data from alternative raw geometric data modalities, such as point clouds. This consideration also aligns with the future work outlined by the DeepCAD method.
The ability to seamlessly convert diverse data representations into CAD models opens new avenues for real-world applications, bridging the gap between different data modalities and expanding the utility of CAD technology. 
This becomes especially critical in scenarios where new variations of a physical object are needed or when repairing a machinery object without access to the corresponding CAD model. This can be particularly challenging in instances where the object predates the digital era of manufacturing.
In these scenarios, the process typically begins with scanning the object using a 3D sensor, which generates a point cloud. Subsequently, the acquired point cloud data needs to be decomposed into a collection of geometric primitives, such as curves or surfaces, to be interpreted by CAD software. The traditional three-step procedure \cite{benkHo2004segmentation} involves converting the point cloud into a mesh, explaining it through parametric surfaces to create a solid (B-Rep), and inferring a CAD program.
Recent advancements in fitting primitives to point clouds \cite{birdal2019generic, li2019supervised, sommer2020primitect} have managed to bypass the initial step of converting point clouds to meshes. Nevertheless, a notable limitation of these methods is their reliance on a finite set of fixed and disjoint primitives, which poses a challenge for convenient shape editing in subsequent steps. 
Addressing this, the recently proposed Point2Cyl \cite{uy2022point2cyl} method frames the problem as an Extrusion Cylinder decomposition task, leveraging a neural network to predict per-point extrusion instances, surface normals, and base/barrel membership. These geometric proxies can then be used to estimate the extrusion parameters through differentiable and closed-form formulations.
In this method, an Extrusion Cylinder is considered a fundamental primitive that signifies an extruded 2D sketch, characterized by parameters such as the extrusion axis, center, sketch, and sketch scale, which can be used to represent 3D CAD models. The terms \textit{base} and \textit{barrel} are utilized to denote specific surfaces of an extrusion cylinder, representing the base/top plane and the side surface of the extrusion cylinder, respectively. 

Point2Cyl leverages PointNet++ \cite{qi2017pointnet} to learn point cloud feature embeddings, which are then passed into two distinct fully connected networks for point cloud segmentation into extrusion cylinder, base/barrel and surface normal prediction. 
The approach is evaluated on Fusion Gallery and DeepCAD datasets, outperforming baselines and showcasing its effectiveness in reconstruction and shape editing.
The code of this method is publicly available.\footnote{point2cyl.github.io} 

Nevertheless, it is important to note that this method is limited in its ability to handle cases where the input data is not noisy or distorted. The task of reconstructing the sharp edges and surfaces in prismatic 3D shapes from noisy point cloud data becomes challenging, given that point clouds inherently offer only an approximate representation of the 3D shape. This challenge is particularly pronounced when dealing with point clouds acquired through low-cost scanners, where any distortions or irregularities present in the shape may be addressed by smoothing during the surface reconstruction process. 
Lambourne et al. \cite{lambourne2022reconstructing} proposed a method concurrently with Point2Cyl, addressing the reconstruction challenge of sharp prismatic shapes when provided with an approximate rounded (smoothed) point cloud. This approach introduces a differentiable pipeline that reconstructs a target shape in terms of voxels while extracting geometric parameters. An autoencoder network is trained to process a signed distance function represented as a voxel grid, obtained through methods converting dense point clouds into signed distance functions \cite{baerentzen2005robust, sanchez2012efficient}. The encoder produces an embedding vector, and the decoders further decompose the shape into 2D profile images and 1D envelope arrays. During inference, CAD data is generated by searching a repository of 2D constrained sketches, extruding them, and combine them via Boolean operations to construct the final CAD model. Evaluation on the ABC dataset demonstrates the method's ability to better approximate target shapes compared to DeepCAD method. 

Concurrent to the two previous works, ComplexGen \cite{guo2022complexgen} has introduced a new approach, ComplexNet, for directly generating B-Rep data from point clouds. This approach reframes the reconstruction task as a holistic detection of geometric primitives and their interconnections, encapsulated within a chain complex structure. ComplexNet, as a neural network architecture, harnesses a sparse CNN for embedding point cloud features and a tri-path Transformer decoder to produce three distinct groups of geometric primitives and their mutual relationships defined as adjacency matrices. Subsequently, a global optimization step refines the predicted probabilistic structure into a definite B-Rep chain complex, considering structural validity constraints and geometric refinements. Extensive experiments on the ABC dataset demonstrate the effectiveness of this approach in generating structurally complete and accurate CAD B-Rep models.


Recently, Ma et al. \cite{ma2024draw} proposed a novel method for reconstructing CAD programs from point clouds using a multimodal diffusion model. Unlike previous methods focused on geometric primitives or B-Rep structures, this approach tokenizes CAD operations and sketches, enabling the generation of editable CAD construction sequences. Through a volume-based diffusion process, it progressively reconstructs human-like design steps, bridging raw 3D data and structured CAD workflows. This method complements prior works by elevating the abstraction from geometry to procedural modeling, making it particularly suitable for interactive reverse engineering tasks.


\subsection{3D CAD Generation from Image}
While CAD generative methods discussed in the previous four categories rely on structured inputs such as sketches, point clouds, or construction sequences, recent advancements have explored generating CAD models directly from unstructured RGB images. These methods attempt to infer structured, editable CAD representations, such as B-reps, from single or multi-view images. This new direction bridges the gap between 2D visual understanding and 3D modeling, enabling applications in reverse engineering, industrial inspection, and conceptual design from photos.

CADDreamer \cite{li2025caddreamer} proposes a pipeline that reconstructs a structured, watertight B-rep CAD model from a single-view RGB image. Unlike traditional 3D reconstruction pipelines that produce meshes or voxels, CADDreamer directly targets CAD-suitable representations by combining primitive-aware diffusion models with geometric optimization.
The core idea of this method is to first generate normal maps and semantic primitive segmentation maps (covering planes, cylinders, cones, spheres, and feature curves) using a multi-view diffusion model. These are then fused across virtual views to build a multi-surface representation of the object. A differentiable renderer then estimates a coarse mesh from the normal maps, while the primitive maps guide segmentation and graph-cut-based patching. The system then performs primitive fitting and topological trimming to reconstruct a valid and editable B-rep with sharp features and semantic structure.
The main advantage of this method is its ability to operate on single-view RGB input, eliminating the need for explicit geometry sensing, while producing compact and editable B-rep models. However, its reconstruction quality is dependent on accurate segmentation and primitive fitting, and the method currently lacks support for parametric design history and user-guided editing.

CADCrafter \cite{chen2025cadcrafter} addresses the same image-to-CAD problem, but instead of generating geometric models directly, it focuses on recovering CAD construction sequences (e.g., sketches and extrusions) that can be imported into CAD software such as Fusion360. This approach reflects real-world design workflows and enables fully editable, parametric outputs.
This method uses a geometry-conditioned latent diffusion transformer that generates a sequence of CAD tokens, which are then decoded into construction commands. The conditioning is achieved by a geometry encoder trained to be texture-invariant, using estimated depth and normal maps derived from the input image. This helps bridge the domain gap between synthetic CAD renders (used during training) and real-world, unconstrained images.
To improve correctness, CADCrafter integrates a code-checking module that compiles and validates the generated CAD commands. 
Furthermore, it introduces a new RealCAD dataset, which contains real-world images paired with command sequences, enabling better generalization.
The main advantage of CADCrafter is its ability to operate on real-world, unconstrained images, supporting both single- and multi-view inputs. It demonstrates strong performance under domain shift, thanks to its geometry-based conditioning, and the use of preference optimization. However, the method is currently limited to sketch-and-extrude workflows and does not yet support more complex or advanced CAD operations.

CADDreamer and CADCrafter demonstrate how deep generative models can automate parametric modeling tasks by interpreting visual content, producing CAD outputs that are both human-readable and machine-executable.

\section{Discussion and Future Works}

Although Geometric Deep Learning (GDL) methods have made remarkable progress in analyzing CAD models and automating the design process at different levels, several challenges remain in this field. 
In this section, some of these challenges are discussed and the potential future research directions for tackling them are proposed.

\noindent
\textbf{Lack of Diversity in Datasets}:
Many CAD models used in real-world product design are protected by intellectual property (IP) rights and cannot be publicly shared. This restriction significantly limits the availability, diversity, and scale of CAD datasets, posing challenges for training machine learning models that rely on large and varied data. As a result, researchers often need to work with synthetic or publicly released datasets, which may not fully capture the complexity of real-world designs.

\noindent
\textbf{Limited Annotated B-Rep Data}: 
Despite the release of several large-scale CAD datasets, which include B-Rep format alongside conventional 3D data formats in recent years, there remains a considerable demand for annotated datasets for supervised learning-based approaches. The datasets annotated for CAD model classification are still limited in size, lacking the diversity and complexity needed for a comprehensive analysis of CAD models.
While there are large-scale annotated datasets for common object CAD shapes, like ShapeNet \cite{chang2015shapenet}, these datasets are only available in mesh format. Consequently, transferring knowledge from common object datasets to mechanical datasets is only viable when using mesh or point cloud data, not B-Rep.
On the other hand, annotating mechanical CAD models is challenging and requires domain knowledge expertise to recognize the type, functionality, or other properties of mechanical objects. Inconsistencies in expert annotations may arise due to different terminologies used for these objects in different industries. 
Similar challenges exist for datasets annotated for CAD segmentation into different faces. As shown in Table \ref{table:datasets}, there are a few annotated datasets in this regard with different types of annotation, such as how different faces are manufactured or the CAD operation used to construct each \textit{face}. 
These datasets lack shared classes in their surface types, making it challenging to transfer learned features from one dataset to another using transfer learning or domain generalization approaches. More extensive datasets, like MFCAD++ extending the MFCAD dataset, are needed in this domain, with different annotation types.
Collecting large-scale CAD datasets that provide various data formats, including mesh, point cloud, and B-Rep, featuring a diverse range of CAD models with various complexities, and annotating them for different tasks, such as CAD classification and segmentation, will significantly contribute to future research and development in this field.

\noindent
\textbf{Analysis on Complex CAD Assemblies}:
Existing works for analyzing CAD assemblies are focusing on how to connect simple parts (solids), predicting the joint coordinates and direction of the connection. A future research direction could be considering how to assemble multiple solids hierarchicaly to make a more complex CAD model, or how to connect two sub-assemblies, each containing multiple parts (solids), to make a bigger assembly. However, the existing datasets are not annotated for such applications, since the joints between different parts are not explicitly specified in the B-Rep data by default. The AutoMate and Fusion 360 Assembly-Joint datasets are annotated specifically for their corresponding methods \cite{jones2021automate, willis2022joinable}, respectively. 
Another open problem in analyzing complex CAD assemblies is to segment each assembly into its building block in different levels (either sub-assembly or part (solid)). Learning how to replace a part in an assembly, while considering all its connections to other parts, its complexity, functionality, materials, etc, with another part in the repository would be a great advantage in automating CAD models customization process. Annotating datasets for this goal is also necessary.

\noindent
\textbf{Representation Learning on B-Rep Data}:
The initial and the most important step in training deep learning models on B-Rep data involves creating numerical feature representations in a format suitable for deep learning architectures. 
Many methods for CAD classification and segmentation, particularly those based on UV-Net, use UV-grid sampling. However, UV-grid features lack permutation invariance. In other words, when the CAD solid is represented as a face adjacency graph, the arrangement of nodes in the graph is crucial for the deep learning model to recognize the graph structure and its similarity to other graphs (or CAD solids), while two graphs with different node arrangements might still be similar together. Hence, exploring various invariances of the B-Rep graph and alternative methods of representing its data in a structured format, while considering the relative orientation of the solid faces and edges, holds promise for future research.

\noindent
\textbf{Unsupervised and Self-Supervised Methods}:
Given the shortage of annotated data for supervised learning in CAD, there is significant potential to enhance self-supervised and/or unsupervised methods, leveraging large-scale datasets like ABC. For instance, UV-Net utilized Graph Contrastive Learning (GCL) for self-supervised learning, and \cite{jones2023self} explored shape rasterization to get the self-supervision from data for training an autoencoder and use the decoder for other supervised tasks. Investigating diverse forms of self-supervision in CAD data for training autoencoders or graph autoencoders would be intriguing. Examining and assessing different graph transformations in GCL approaches to understand the structure of B-Rep graph locally and globally, maintaining various invariances, is a crucial avenue for future research. Additionally, focusing on Variational Autoencoders for direct synthesis on B-Rep data, generating meaningful new shapes from the same distribution as existing shapes, would enhance similarity analysis and CAD retrieval.

\noindent
\textbf{CAD Generation and ‌B-Rep Synthesis}:
The current generative methods for CAD commands often focus on a limited set of operations, like \textit{Sketch} and \textit{Extrude}, restricting the complexity and diversity of the resulting CAD models. Notably, there is no assurance in certain methods, such as DeepCAD \cite{wu2021deepcad}, that all generated CAD command sequences will yield topologically valid CAD models, particularly in cases of complex models with long command sequences. Therefore, an avenue for future exploration involves expanding generative methods to encompass a broader range of CAD operations, such as \textit{fillet} and \textit{chamfer}, allowing the generation of command sequences for more complex CAD shapes. Additionally, advancements in methods like SolidGen \cite{solidgen23} and CADParser \cite{zhou2023cadparser}, which directly focus on synthesizing B-Rep data without relying on CAD command sequences, mark a promising direction. Despite these advancements, there is still room for creativity and improvement in this area.
Another direction for future research involves extending generative methods to produce CAD construction operations or B-Rep synthesis from other 3D data formats, like mesh, voxel and point cloud. Recent progress in converting point clouds to CAD models, described in Section \ref{sec:cad2pc}, opens possibilities for transferring knowledge across different data domains. Moreover, generating 3D CAD models from noisy scans or hand-drawn sketches is in high demand in CAD tools and presents significant potential for improvement. Although some methods have been introduced in this domain, the lack of a paired sketch and B-Rep dataset makes training supervised learning methods currently infeasible. For example, the sketches of SketchGraphs and 3D models in ABC datasets are both collected from Onshape repository. However, the 2D and 3D models in these two datasets are not paired. Collecting such a paired dataset would greatly aid future research. 

\noindent
\textbf{Manufacturing-Aware and Simulation-Ready Geometry Understanding}
While significant progress has been made in GDL for tasks such as shape classification, generative design, and parametric modeling, downstream engineering applications, including Design for Manufacturing (DFM), simulation-ready B-Rep healing, and engineering constraint solving, remain relatively underexplored. These tasks are critical for bridging the gap between CAD design and real-world deployment, yet the current body of GDL research often overlooks them.
Most existing works rely on traditional rule-based or physics-driven methods to evaluate manufacturability or prepare geometry for simulation. However, the structured, semantic-rich nature of B-Rep data, combined with large-scale industrial repositories, presents an opportunity for deep learning models to learn manufacturability heuristics, predict constraint violations, or even autonomously repair geometric flaws that prevent meshing or analysis.
Future research can build on the foundations of B-Rep–based GDL models by targeting these high-impact engineering tasks. For example, learning-based DFM checkers could provide real-time feedback during the design process; learning-based healing models could automatically fix non-manifold or noisy CAD models for simulation; and constraint-predictive models could support reverse engineering of legacy geometry or enforce design intent in generative systems. Developing such systems requires not only algorithmic advances, but also benchmark datasets, ground truth annotations, and closer collaboration with manufacturing and simulation experts.


\noindent
\textbf{Reproducibility}:
In this domain, a significant challenge lies in reproducing and comparing the experimental results of different methodologies. The absence of large-scale annotated benchmark datasets leads each method proposed for machine learning-based CAD analysis to either introduce a new annotated CAD dataset tailored for the specific task or modify and annotate a portion of a large-scale dataset for evaluation. Supervised methods trained on these smaller datasets often exhibit high performance, leaving little room for improvement. Moreover, since each method is benchmarked on a dataset adapted for its specific task, comparing results becomes complex. For instance, when analyzing the results of different methods evaluated on a subset of the Fusion 360 segmentation dataset in \cite{jones2023self}, it is unclear which part of the dataset each method used for evaluation.
Reproducing the results of different methods and conducting performance comparisons become especially challenging when researchers do not release their preprocessed data or do not provide clear preprocessing instructions. Another hurdle in replicating research results is the reliance of codes on CAD kernels that are not open-source. For instance, replicating the results of AutoMate \cite{jones2021automate} requires installing the Parasolid kernel, which is typically accessible only to industrial developers and companies. This limited availability makes it challenging for independent academic researchers to utilize and build upon such research works. While codes dependent on the open-source OpenCASCADE kernel are more accessible, many methods in this field either use constraint solvers through CAD tools during training or are in a way reliant on a CAD tool kernel, necessitating access and licensing for at least one CAD software.
To address these issues, it is highly recommended that researchers provide comprehensive documentation for their released codes, detailing the data preprocessing setup and offering sufficient information on the experimental setting and code dependencies. This transparency can significantly facilitate future research efforts in the field.

In summary, despite recent advances in GDL for CAD analysis and generation, several key challenges remain. These include limited dataset diversity, lack of annotated B-Rep data, difficulties in handling complex assemblies, and issues with reproducibility. Many methods rely on proprietary software or private datasets, making it difficult to reproduce results or establish fair benchmarks, and the use of non-overlapping datasets across tasks further limits direct comparability. These limitations also present substantial opportunities for future research, such as developing scalable representation learning, advancing self-supervised approaches, designing richer generative models, and bridging CAD with manufacturing and simulation tasks. Addressing these directions will be essential for enabling robust, intelligent, and automation-ready CAD systems.

\section{Conclusion}
Developing and training Geometric Deep Learning (GDL) models to learn and reason about CAD designs holds the promise of revolutionizing design workflows, bringing in greater efficiency. However, extending machine learning-based methods to complex parametric data, like B-Rep, poses a crucial and challenging task. 
This survey offers a comprehensive review of GDL methods tailored for CAD data analysis. It presents detailed explanation, comparisons, and summaries within two primary categories: 1) CAD representation learning, encompassing supervised and self-supervised methods designed for CAD classification, segmentation, assembly, and retrieval, and 2) CAD generation, involving generative methods for 2D and 3D CAD construction. Additionally, benchmark datasets and open-source codes are introduced. The survey concludes by discussing the challenges inherent in this rapidly evolving field and proposes potential avenues for future research.

\section*{Acknowledgement}
The research leading to the results of this paper received funding from the Thomas B. Thriges Foundation and the Industriens Foundation as part of the AI Supported Modular Design and Implementation project.

\bibliographystyle{IEEEtran}
\bibliography{references}

\end{document}